\documentclass[12pt]{report}
\usepackage[T1]{fontenc}
\usepackage{helvet}
\usepackage{geometry}
\usepackage{graphics}

\makeatletter

\newcommand{\LyX}{L\kern-.1667em\lower.25em\hbox{Y}\kern-.125emX\spacefactor1000}

\usepackage[T1]{fontenc}


\begin{document}

\title{Drag Reduction over Dolphin Skin via the Pondermotive Forcing of Vortex Filaments}

\author{A. G. Lisi\thanks{
Gar@Lisi.com
}}

\date{May 4, 1999}

\maketitle
\begin{abstract}
The skin of \textit{Tursiops Truncatus} is corrugated with small, quasi-periodic
ridges running circumferentially about the torso. These ridges extend into the
turbulent boundary layer and affect the evolution of coherent structures. The
development and evolution of coherent structures over a surface is described
by the formation and dynamics of vortex filaments. The dynamics of these filaments
over a flat, non-ridged surface is determined analytically, as well as through
numerical simulation, and found to agree with the observations of coherent structures
in the turbulent boundary layer. The calculation of the linearized dynamics
of the vortex filament, successful for the dynamics of a filament over a flat
surface, is extended and applied to a vortex filament propagating over a periodically
ridged surface. The surface ridges induce a rapid parametric forcing of the
vortex filament, and alter the filament dynamics significantly. A consideration
of the contribution of vortex filament induced flow to energy transport indicates
that the behavior of the filament induced by the ridges can directly reduce
surface drag by up to \( 8\% \). The size, shape, and distribution of cutaneous
ridges for \textit{Tursiops Truncatus} is found to be optimally configured to
affect the filament dynamics and reduce surface drag for swimming velocities
consistent with observation.
\end{abstract}
\tableofcontents

\listoffigures

\chapter{Introduction}

\section{Dolphins}

A dolphin gliding through water with grace and speed presents both a beautiful
performance of nature and a persistent enigma to fluid dynamicists. Traditional
models of fluid flow applied to a swimming dolphin suggest this mammal is capable
of aberrantly large powers of muscular exertion. However, studies of dolphin
energy expenditure show that a dolphin has a metabolism similar to that of other
mammals \cite{twill}, implying that these animals, rather then having unusual
muscular capability, may posses some mechanism that allows them to slip through
water with unusually little effort. Sir John Gray was the first to hypothesize
that dolphins may posses an ability to impede the development of turbulence
over the skin surface and hence reduce the surface drag of a swimming dolphin
\cite{jgray}. A method of reducing boundary layer turbulence and surface drag
would be extremely useful if it could be implemented efficiently, and the actualization
of such a method has posed a theoretical and experimental challenge to researchers.

\subsection{Skin pliability and active control response}

Some researchers have attempted to reproduce the dolphin's conjectured drag
reduction by studying -- and attempting to duplicate -- the pliable nature of
dolphin skin \cite{mkramer}. It was thought that the surface drag induced by
a turbulent boundary layer could be reduced if the surface was able to yield
to the microscopic disturbances produced by turbulent eddies. Unfortunately,
this approach was not significantly successful. More recently, a large effort
has been extended towards exploring the possibility that active control of surface
shape in response to the forces from the turbulent boundary layer may be able
to reduce turbulent development. The proposal that this sort of direct active
response is implemented by dolphins is ostensibly supported by the existence
of systems of nerves and cutaneous muscle that appear capable of actively controlling
the skin surface. However, detailed measurement of dolphins' skin reaction to
microvibratory stimuli indicate a minimum response time on the order of \( 10^{-2}s \)
\cite{sr}, whereas the timescale associated with the passage of coherent structures
over the skin is on the order of \( 10^{-4}s \) \cite{habar1}. This indicates
that a dolphin's skin is not able to respond to the passage of individual coherent
structures.

\subsection{Cutaneous ridges}

\begin{figure}
{\centering \resizebox*{5.5in}{!}{\includegraphics{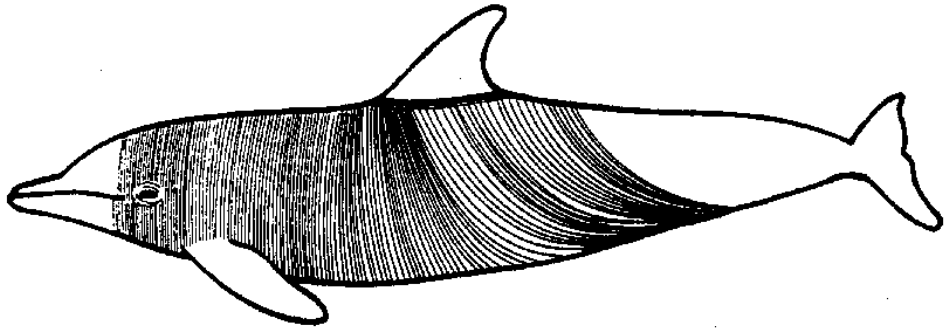}} \par}

\caption{\label{fig: ridges}A sketch showing the distribution and orientation of cutaneous
ridges over the body of a dolphin (From Shoemaker and Ridgway \cite{sr}).}
\end{figure}

It is only recently that Shoemaker and Ridgway documented a feature of odontocete
skin that had been virtually unknown previous to their work; namely, the existence
of cutaneous ridges covering most of the body surface \cite{pshoe}. These corrugations,
with furrows running circumferentially around the whales, are reported to be
\( h\simeq 10-60\mu m \) in height when the animal is at rest, with a separation
of \( L\simeq 0.4-1.7mm \), and vary slightly in scale over the dolphin's body.
The ridges are absent from the dolphin's snout and from much of the head --
regions associated with laminar flow, as well as from the flippers and tail,
(Figure \ref{fig: ridges}). The fact that the corrugations exist only in regions
of steady turbulent flow is a strong indication that these ridges are associated
with turbulent development. The absence of the corrugations from the flippers
and tail indicates that the ridges may not be useful in these regions which
are exposed to large velocity and pressure variations.

\begin{figure}
{\centering \resizebox*{5.5in}{!}{\includegraphics{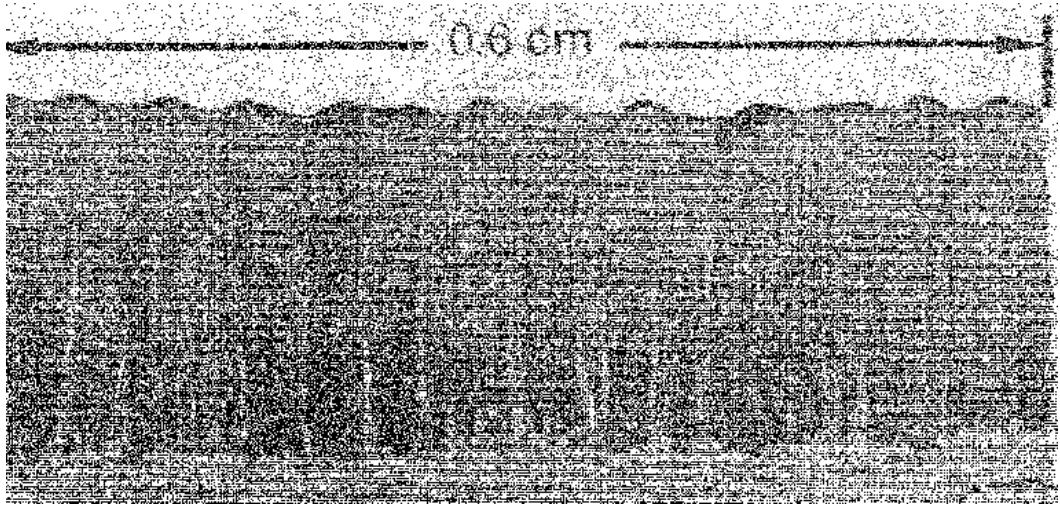}} \par}

\caption{\label{fig: skin}A photo-micrograph of a cross section of dolphin skin, showing
the cutaneous ridges and underlying dermal structure (From Shoemaker and Ridgway
\cite{sr}).}
\end{figure}

Microscopic studies of  sections of skin over the dolphin's body show the capillary
and muscular structure to be correlated with cutaneous ridge spacing. This anatomical
analysis is consistent with the possibility that a dolphin may actively tune
the amplitude of surface corrugations to correspond with swimming velocity.
Although the measured surface ridge height is small for a dolphin at rest, producing
a ridge height to length ratio of \( \frac{h}{L}\simeq 0.025 \), a photo-micrograph
of a section of dolphin skin, (Figure \ref{fig: skin}), displays a much larger
ratio, \( \frac{h}{L}\simeq 0.2 \). This indicates that a live dolphin may
be reducing the ridge height to the smaller value while at rest by using cutaneous
muscle, while the ``relaxed'' ridge height ratio, indicative of the available
range of ridge heights, is much larger. This hypothesis is further supported
by reports from trainers at Sea World that the surface ridges become more pronounced
with greater swimming speed \cite{pshoe}.

The existence of regular corrugations with tunable amplitude covering the bodies
of these efficient swimmers, in regions associated with steady turbulent boundary
layer flow, presents a challenging new mystery. This challenge is met in this
dissertation, which presents a theoretical model for and an analysis of the
effect of such a wavy boundary on turbulent development.

\subsection{Fluid flow development over dolphin skin}

Allthough cutaneous ridges were observed and measured by Shoemaker and Ridgway
in seven species of toothed whales (odontocetes), ranging in body length from
two to eight meters \cite{pshoe}, no correlation between body size and corrugation
scale is visible in the data. It may be possible, through future study, to establish
a correlation between corrugation scale and swimming speed. However, detailed
information regarding the swimming speeds of these various species is not currently
available. Hence, a correlation between swimming speed and corrugation size
is not presently possible to establish. The best observational data currently
exists for the bottlenosed dolphin (\textit{Tursiops Truncatus}), and this species
is taken as representative for this analysis.  

Adult \textit{Tursiops} vary in size from two to four meters, mass approximately
\( 200kg \), and have been observed to be capable of bursts of swimming speed
up to \( 50\frac{km}{hr} \), though constant cruising speed is reported to
be roughly \( U\simeq 10\frac{km}{hr} \) \cite{dri}.

Since the boundary layer thickness at the location where the transition to turbulence
occurs is small, approximately 
\begin{equation}
\label{dispthick}
\delta _{*}^{c}\simeq 420\frac{\nu }{U}=420\frac{10^{-6}\frac{m^{2}}{s}}{10\frac{km}{hr}}=.15mm
\end{equation}
at cruising speed, the circumferential curvature of the dolphin's body is not
significant. Thus, a laterally flat dolphin may be assumed for the purpose of
studying boundary layer turbulence. The transition to turbulence occurs for
flow over a flat plate at a calculated position of 
\begin{equation}
\label{transx}
x^{c}\simeq 6\times 10^{4}\frac{\nu }{U}=6\times 10^{4}\frac{10^{-6}\frac{m^{2}}{s}}{10\frac{km}{hr}}=2.2cm
\end{equation}
from the edge of the plate. However, transition to turbulence in flow over a
dolphin is significantly delayed by the pressure gradient induced by the longitudinal
curvature of the dolphin's head, and is observed via bioluminescent excitation
to begin approximately \( 40cm \) from the snout \cite{jrohr}.

Allthough transition to turbulence in flow over a dolphin's curved head is delayed
by the favorable pressure gradient, once initiated, turbulent development and
evolution is assumed to be of the same character as turbulent development over
a flat plate with no pressure gradient.

\chapter{The boundary layer}

A consideration of the effect of surface corrugations on turbulent flow development
starts with understanding the evolution of the boundary layer over a flat surface.

The flow of water, with density \( \rho \simeq 10^{3}\frac{kg}{m^{3}} \) and
viscosity \( \mu \simeq 10^{-3}\frac{kg}{m\, s} \), is described by the Navier-Stokes
equation,
\begin{equation}
\label{navierstokes}
d_{t}\overrightarrow{u}=\partial _{t}\overrightarrow{u}+\left( \overrightarrow{u}\cdot \overrightarrow{\nabla }\right) \overrightarrow{u}=-\frac{1}{\rho }\overrightarrow{\nabla }p+\nu \nabla ^{2}\overrightarrow{u}
\end{equation}
in which \( \nu =\frac{\mu }{\rho }\simeq 10^{-6}\frac{m^{2}}{s} \) is the
kinematic viscosity, \( d_{t} \) is the advective derivative, and the fluid
velocity field, \( \overrightarrow{u}=u_{x}\widehat{x}+u_{y}\widehat{y}+u_{z}\widehat{z} \),
is restricted to be incompressible,
\begin{equation}
\label{incompress}
\overrightarrow{\nabla }\cdot \overrightarrow{u}=0
\end{equation}
Incompressibility may be assumed, provided fluid velocities remain much less
than the speed of sound in the fluid. All velocities are described with respect
to the rest frame of the boundary surface. Therefore, the skin of a dolphin
swimming at speed \( U \) through stationary fluid is modeled as a stationary
flat plate immersed in fluid with background (free stream) velocity of \( \lim _{y\rightarrow \infty }u_{x}=U \)
in the \( \widehat{x} \) (streamwise) direction. The plate is oriented such
that \( \widehat{y} \) is normal to the plate surface, which lies in the plane
\( y=0 \), with the leading edge of the plate located at \( x=0 \). The \( z \)
coordinate runs along the span of the plate, with arbitrary origin.

\section{The laminar boundary layer}

\begin{figure}
{\centering \resizebox*{3.5in}{!}{\includegraphics{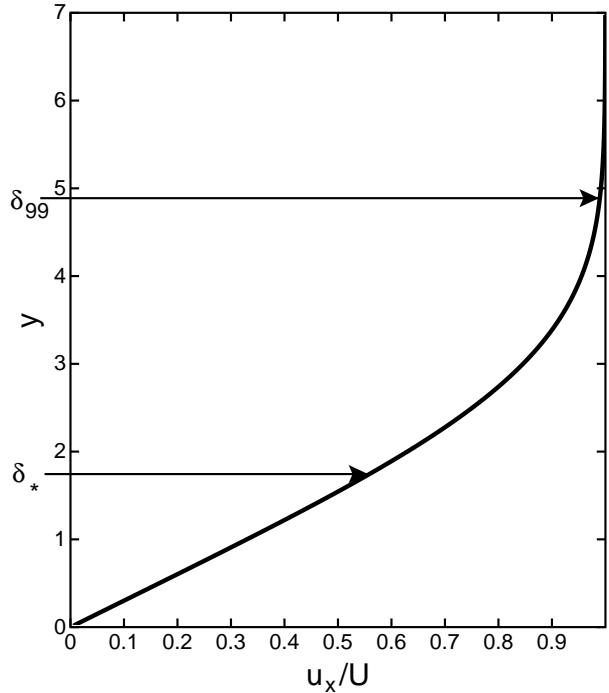}} \par}

\caption{\label{fig: blas}The horizontal velocity profile, \protect\( \frac{u_{x}(y)}{U}\protect \),
of the laminar boundary layer. The vertical scale is in units of \protect\( \sqrt{\frac{\nu x}{U}}\protect \).
The heights of the displacement thickness, \protect\( \delta _{*}\protect \),
and \protect\( 99\%\protect \) thickness, \protect\( \delta _{99}\protect \),
are shown.}
\end{figure}

Viscosity plays the primary role in boundary layer evolution as the irrotational
fluid first encounters the leading edge of the flat plate. The fluid at the
plate surface is attached to the surface via molecular interactions and must
satisfy the ``no slip'' boundary condition,
\begin{equation}
\label{noslip}
\overrightarrow{u}(x>0,\, y=0)=0
\end{equation}
The equations of motion, (\ref{navierstokes}), for the flow may be solved numerically
or approximately to obtain the laminar boundary layer profile, \( u_{x}(x,y) \),
(Figure \ref{fig: blas}). The profile may be characterized by several descriptors
of boundary layer thickness. The \( 99\% \) thickness, \( \delta _{99} \),
is the most common descriptor, and is defined arbitrarily as the height at which
\( u_{x}(y=\delta _{99})=.99U \). Another less arbitrary descriptor, the displacement
thickness, \( \delta _{*} \), is defined to be equal to the distance the wall
would have to be displaced into the fluid such that the mass flux of fluid over
the boundary would be the same as that for fluid flowing over the wall ``free
slip'' with the free stream velocity \( U \),
\[
\delta _{*}=\int _{0}^{\infty }\left( 1-\frac{u_{x}}{U}\right) dy\]
and may be considered the middle of the boundary layer.  

The boundary layer fluid induces a shear stress of
\[
\tau _{0}=\left. \mu \frac{\partial u_{x}}{\partial y}\right| _{y=0}\]
on the wall in the \( \widehat{x} \) direction. This stress produces the drag
force per unit area on a body traveling through the fluid. The shear stress
at the surface provides a natural scale for velocities and distances near the
wall for both laminar and turbulent flow. The friction velocity is defined as
\[
u_{*}=\sqrt{\frac{\tau _{0}}{\rho }}\]
and the length scale, one ``wall unit,'' is defined as 
\begin{equation}
\label{wu}
y_{*}=1wu=\frac{\nu }{u_{*}}
\end{equation}

The boundary layer thickness grows as the fluid progresses over the body. For
flow over a flat plate, the laminar boundary layer displacement thickness grows
as
\[
\delta _{*}\simeq 1.72\sqrt{\frac{\nu x}{U}}\]
The corresponding boundary layer Reynolds number, \( Re_{*}=\frac{U\delta _{*}}{\nu } \),
increases with the boundary layer thickness until it reaches a critical value
of \( Re_{*}^{c}\simeq 420 \), at which point the laminar flow becomes unstable.
This occurs at the corresponding critical boundary layer thickness of
\[
\delta _{*}^{c}\simeq 420\frac{\nu }{U}\]
and, for a flat plate, at the position
\[
x^{c}\simeq 6\times 10^{4}\frac{\nu }{U}\]
when the boundary layer becomes unstable to the growth of TS waves. The Reynolds
number describing the transition point is also often give as \( Re_{x}^{c}=\frac{Ux^{c}}{\nu } \),
corresponding to the distance from the plate edge at which transition occurs.

Because the shear stress, \( \tau _{0} \), decreases as the boundary layer
thickness, \( \delta _{99} \), grows, the wall units of the laminar boundary
layer do not grow as quickly in \( x \) as the boundary layer thickness. The
wall units for the laminar profile grow as
\[
y_{*}=\frac{\nu }{u_{*}}\simeq 1.74\left( \frac{\nu x}{U}\right) ^{\frac{1}{4}}\left( \frac{\nu }{U}\right) ^{\frac{1}{2}}\]
At the transition point these wall units are \( y^{c}_{*}\simeq 27\frac{\nu }{U} \).

\section{The boundary layer TS instability}

The linear stability analysis of laminar boundary layer flow over a flat plate
was first calculated by Tollmien in 1929, Schlichting in 1933, and confirmed
experimentally by Schubauer and Skramstadt in 1947. The stability analysis proceeds
by considering an infinitesimal, \( z \) independent perturbation, \( \overrightarrow{u_{1}} \),
to the laminar velocity field, \( \overrightarrow{u_{0}} \), of the form
\[
\overrightarrow{u}=\overrightarrow{u_{0}}+\overrightarrow{u_{1}}(y)e^{i(kx-\omega t)}\]
and solving the resulting eigenvalue problem for the wavenumber, \( k \), frequency,
\( \omega  \), and perturbing profile, \( \overrightarrow{u_{1}}(y) \). For
a laminar velocity profile, \( \overrightarrow{u_{0}} \), determined from a
solution of the Navier-Stokes equations, the imaginary component of \( \omega  \)
first becomes positive when the boundary layer has grown to the thickness \( \delta _{*}^{c}\simeq 420\frac{\nu }{U} \),
implying the onset of instability. The frequency and wavelength of the unstable
TS waves at this onset of instability are found to be \( \Re (\omega ^{c})\simeq .15\frac{U}{\delta _{*}^{c}} \)
and \( \lambda _{x}^{c}=\frac{2\pi }{k^{c}}\simeq 17\delta _{*}^{c} \). The
TS waves grow very rapidly beyond the transition point, and the boundary layer
takes on a qualitatively different form. The new flow, now with large scale
periodic variations in \( x \), becomes unstable to a three-dimensional instability
with periodic variations in \( z \). The three-dimensional instability, best
understood through a study of vorticity dynamics, evolves directly to turbulent
flow.

\section{The evolution of boundary layer vorticity }

The evolution of the boundary layer may be more deeply understood via a description
of the dynamics of vorticity. The vorticity field of the fluid is defined as
the curl of the velocity,
\[
\overrightarrow{\omega }=\overrightarrow{\nabla }\times \overrightarrow{u}\]
Taking the curl of the Navier-Stokes equation, (\ref{navierstokes}), and using
a vector identity produces the vorticity equation for the incompressible fluid,
\begin{equation}
\label{vorticity}
d_{t}\overrightarrow{\omega }=\partial _{t}\overrightarrow{\omega }+\left( \overrightarrow{u}\cdot \overrightarrow{\nabla }\right) \overrightarrow{\omega }=\left( \overrightarrow{\omega }\cdot \overrightarrow{\nabla }\right) \overrightarrow{u}+\nu \nabla ^{2}\overrightarrow{\omega }
\end{equation}
The laminar fluid flow over the flat plate is described simply as the creation
of concentrated vorticity at the leading edge and its diffusion, advection,
and stretching downstream.

\subsection{Formation of vortex filaments}

The advected vorticity initially coats the flat plate with a smooth layer of
vorticity. The entire laminar boundary layer may be considered a vortex sheet
of constant strength, with sheet thickness increasing with \( x \). From this
point of view, the TS instability of the boundary layer is seen as similar to
the Kelvin-Helmholtz instability of a shear layer.

With the onset of the TS instability, the smooth vorticity of the sheet condenses
into areas of higher vorticity separated periodically in \( x \). These concentrations
of vorticity then continue via nonlinear evolution to wrap and condense into
spanwise oriented vortex filaments, evolution similar to the nonlinear development
of a shear layer \cite{psaff}. This process of nonlinear development into vortex
filaments is highly influenced by small irregularities in the initial fluid
velocity field which produce significant variations from perfectly uniform vortex
filaments. Although the shape and distribution may deviate significantly from
the ideal, these filaments and their evolution represent the dominant structure
of the boundary layer after the transition point, and play the central role
in further boundary layer development.

\subsection{Three-dimensional vortex filament instability}

The initial vortex filaments, evolving from the vortex sheet after the TS instability,
are separated by the distance \( \lambda _{x}^{c}\simeq 17\delta _{*}^{c} \)
in \( x \), and are located at the initial height \( y_{0} \) above the surface.
A calculation of the three dimensional instability will show that this height
must be equal to \( y_{0}\simeq 6wu \). Since \( \lambda _{x}^{c}\gg y_{0} \),
the effect of initial neighboring filaments on evolution is small and may be
ignored \cite{habar2}.

\begin{figure}
{\centering \resizebox*{4in}{!}{\includegraphics{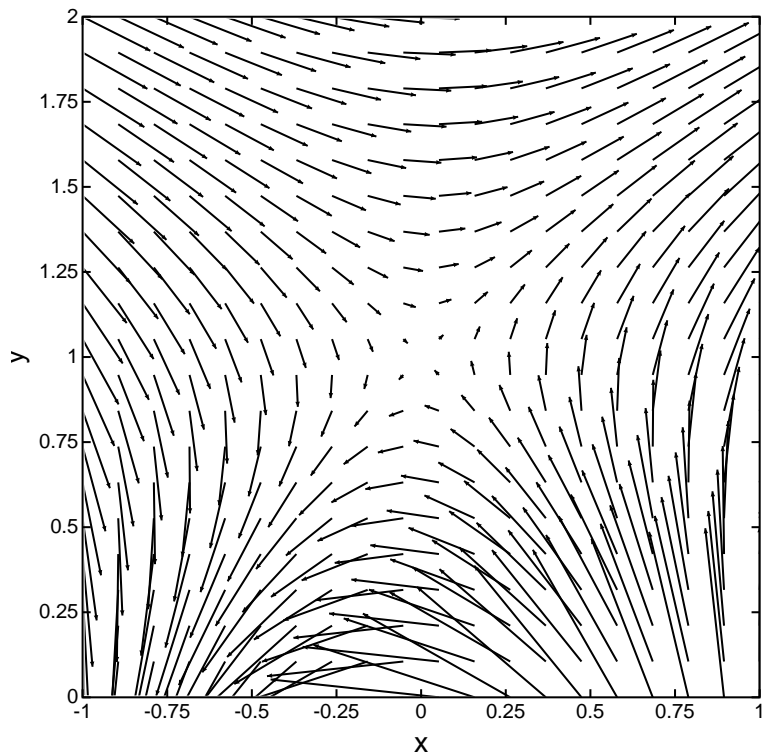}} \par}

\caption{\label{fig:velocity}The velocity field induced by the image vortex, relative
to the motion of a straight vortex filament located at \protect\( x=0\protect \),
\protect\( y=1\protect \). The image vortex is located at \protect\( x=0\protect \),
\protect\( y=-1\protect \). The self-induced velocity field and background
flow field are not included.}
\end{figure}

A straight vortex filament in the presence of a boundary is unstable due to
the velocity field induced by its interaction with the surface. The filament
velocity field is calculated by considering it to arise from an image vortex
beneath the flat surface. A plot of the image induced velocity field surrounding
the filament, relative to the motion of the filament, (Figure \ref{fig:velocity}),
shows the nature and direction of the stretching flow. 

\begin{figure}
{\centering \resizebox*{5.5in}{!}{\includegraphics{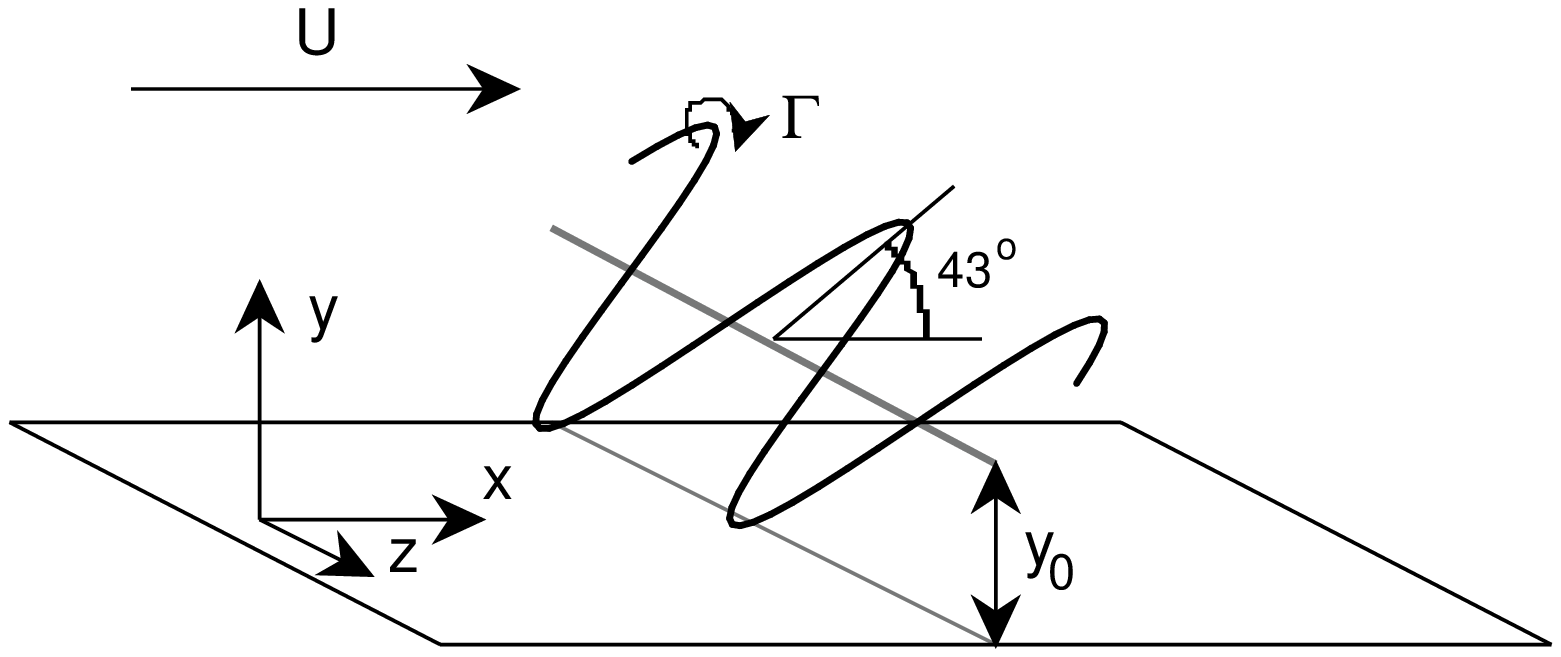}} \par}

\caption{\label{fig:filament}A plot of the maximally unstable deformation mode of a
vortex filament over a flat surface.}
\end{figure}

The filament is linearly unstable to deformations in this stretching plane,
oriented at an angle of approximately \( 45^{o} \) to the boundary surface.
This three dimensional vortex filament instability is treated in detail in Chapter
\ref{sec:flatboundary}, and the effects of a wavy boundary are considered in
Chapter \ref{sec: wavy}. The filament over a flat boundary is found to be maximally
unstable to deformations of wavelength \( \lambda ^{c}_{z}\simeq 17y_{o} \),
and is sketched in Figure \ref{fig:filament}.

A numerical investigation confirms that a filament with a small initial perturbation
rapidly deforms into a sinusoid of wavelength \( \lambda ^{c}_{z} \) inclined
at an angle of approximately \( 43^{o} \). The expanding sinusoidal filament
then leaves the realm of linear evolution as the bottom of the filament nears
the surface and adheres to it, forming streamwise oriented vortex ``legs,''
while the top of the filament continues to rise with linear rather then exponential
growth as it advects downstream. In this manner a ``horseshoe'' or ``hairpin''
vortex is formed. Although the existence of hairpin vortices was postulated
by Theodorsen in 1952 \cite{ttheod}, and has been examined in great detail
in experiments by Head and Bandyopadhyay \cite{mhead} and others \cite{srobins},
the vortex filament instability mechanism, the Crow instability \cite{scrow},
leading to the formation of these hairpin vortices, has only now been established
\cite{habar2}. The linear Crow instability of a vortex filament pair was initially
calculated, and confirmed numerically \cite{dmoore}, for the case of two aircraft
trailing vortices. The same analysis applies for the near wall filament, and
is described in Chapter \ref{sec:flatboundary}, with the second of the vortex
pair replaced by the image filament.

Recent experimental investigations indicate that autogeneration occurs in the
further evolution of these vortex filaments \cite{jzhou}. In this process,
the legs of the hairpin vortex pinch together and form another filament, which
then grows via the same instability, to align itself slightly behind the original.
This process repeats and produces a long chain of hairpin vortices in line behind
each original hairpin. These trains of tightly packed hairpin vortices, ``hairpin
packets,'' are the dominant structures in the turbulent boundary layer.

\section{The turbulent boundary layer}

The fully turbulent boundary layer over a flat plate is a tumultuous collection
of hairpin vortex packets and filament remnants. Since boundary layer evolution
is highly nonlinear and sensitively dependent on initial conditions, any simple
model, such as the one described with the idealized vortex filament picture,
can only represent the dominant processes of the dynamics and cannot capture
the fine structure of small scale motions. Rather, a statistical analysis of
fluid flow is employed to understand the processes of energy transport and induced
surface drag. The effects of coherent structures, such as collections of hairpin
vortices, may then be considered on the basis of how they effect the mean properties
of the flow.

\subsection{Reynolds decomposition}

The mean velocity field in the turbulent boundary layer is found by taking the
infinite time average of the velocity field
\[
\overline{\overrightarrow{u}}=\lim _{t_{0}\rightarrow \infty }\frac{1}{t_{0}}\int _{0}^{t_{0}}\overrightarrow{u}dt\]
The Reynolds decomposition is performed by decomposing the variables into the
time average quantity and the remaining fluctuating quantity, \( \overrightarrow{u'} \),
\[
\overrightarrow{u}=\overline{\overrightarrow{u}}+\overrightarrow{u'}\]
This decomposition allows the steady, time independent, mean flow to be considered
separately from the more complex underlying turbulent flow. The mean equations
of motion are obtained by taking the time average of the incompressibility condition,
(\ref{incompress}), and Navier-Stokes equation, (\ref{navierstokes}), to get,
in component notation,
\begin{eqnarray}
\partial _{i}\overline{u_{i}} & = & 0\\
d_{t}\overline{u_{i}}=\overline{u_{j}}\partial _{j}\overline{u_{i}} & = & -\frac{1}{\rho }\partial _{i}\overline{p}+\nu \partial _{j}\partial _{j}\overline{u_{i}}-\partial _{j}\left( \overline{u_{i}'u_{j}'}\right) \label{meanmom} 
\end{eqnarray}
The last term, \( -\partial _{j}\left( \overline{u_{i}'u_{j}'}\right)  \),
the Reynolds stress term, represents the effect of the turbulent eddies on the
mean flow. 

\begin{figure}
{\centering \resizebox*{3.5in}{!}{\includegraphics{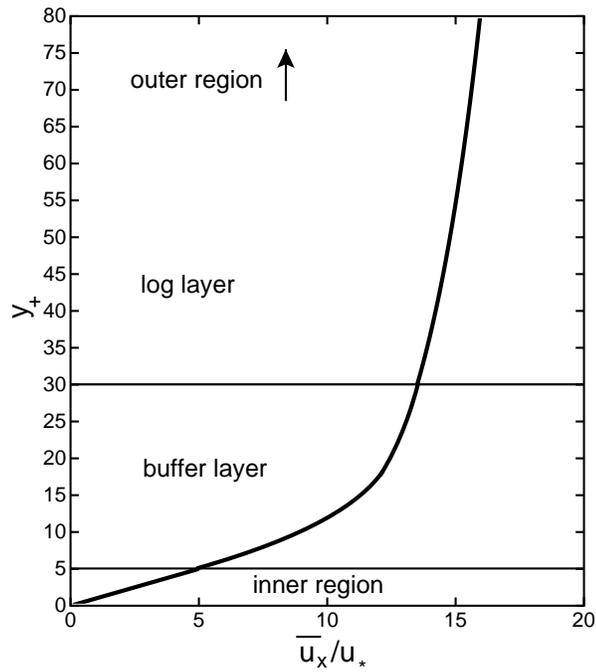}} \par}

\caption{\label{fig:tprof}The turbulent boundary layer mean velocity, \protect\( \overline{u_{x}}(y)\protect \),
showing the various regions of the boundary layer. The mean velocity, along
the horizontal axis, is shown as a function of the height above the surface
in wall units, \protect\( y_{+}=\frac{y}{y_{*}}\protect \), along the vertical
axis.}
\end{figure}

The mean flow of the boundary layer develops a profile similar to the laminar
flow profile. A standard model of the mean velocity profile for turbulent boundary
layer flow has emerged from many decades of experimental research. The profile,
\( \overline{u_{x}} \), as a function of \( y \), breaks into several regions.
The ``inner region'' extends to approximately \( 5wu \), and is fitted by
the curve \( \overline{u_{x}}(y<5wu)\simeq u_{*}\frac{y}{y_{*}} \), with \( \frac{y}{y_{*}} \)
equal to the non-dimensionalized distance from the surface in wall units as
defined by (\ref{wu}). The ``log layer'' begins at approximately \( 30wu \),
and is fitted by the curve \( \overline{u_{x}}\simeq u_{*}\left( \frac{1}{.4}\ln (\frac{y}{y_{*}})+5\right)  \).
These curves are matched to each other in the ``buffer layer'' between \( 5 \)
and \( 30wu \). The log layer extends out into the ``outer region,'' where
it asymptotically approaches the free stream flow, \( U \), (Figure \ref{fig:tprof}). The mean boundary layer thickness grows linearly with \( x \) as \( \overline{\delta _{99}}\simeq .4\frac{u_{*}}{U}x \),
faster then the square root dependence of the laminar boundary layer. 

Unlike the laminar flow case, in which the primary energy loss is due to viscosity,
the mean flow of the turbulent boundary layer is influenced more strongly by
the energy loss to the underlying turbulent flow due to the Reynolds stress.
In regions far from the boundary the effects of viscosity on the mean equations
of motion are negligible, and the viscosity term may be ignored.

\subsection{Energy flow}

The mean energy transport equation is obtained by multiplying (\ref{meanmom})
by \( \overline{u_{i}} \) to get 
\[
d_{t}\left( \frac{1}{2}\overline{u_{i}}\overline{u_{i}}\right) =\partial _{i}\left( -\frac{1}{\rho }\overline{p}\overline{u_{i}}+2\nu \overline{u_{j}}E_{ij}-\overline{u_{i}'u_{j}'}\overline{u_{j}}\right) -2\nu E_{ij}E_{ij}+\overline{u_{i}'u_{j}'}\partial _{j}\overline{u_{i}}\]
in which the mean strain rate is \( E_{ij}=\frac{1}{2}\left( \partial _{i}\overline{u_{j}}+\partial _{j}\overline{u_{i}}\right)  \).
This equation is best understood by considering it a description of the mean
energy balance of an infinitesimal fluid element. The change in kinetic energy
of the element, \( d_{t}\left( \frac{1}{2}\overline{u_{i}}\overline{u_{i}}\right)  \),
is equal to the energy transported across its boundary (the divergence term),
plus the loss to viscosity, \( -2\nu E_{ij}E_{ij} \), plus the loss to turbulence,
\( \overline{u_{i}'u_{j}'}\partial _{j}\overline{u_{i}} \).

The most significant mean strain in the boundary layer is \( \partial _{y}\overline{u_{x}} \).
Hence, the dominant energy transport from mean to turbulent motion is due to
the term \( \overline{u_{x}'u_{y}'}\partial _{y}\overline{u_{x}} \), and large
energy losses to turbulence are associated with \( \overline{u_{x}'u_{y}'}<0 \)
events in a region of large mean shear. This may be understood physically as
the creation of turbulent flow by the lifting of low speed fluid near the boundary
layer up and against the higher speed fluid in the upper boundary layer, a ``low
speed streak'' and ``turbulent burst,'' and by the converse, a ``sweep,''
in which high speed fluid descends into the lower boundary layer \cite{skline}.
This intermittent turbulent bursting and sweeping phenomenon is experimentally
observed in the turbulent boundary layer, and measured to account for the majority,
\( 60-80\% \) \cite{agrass}\cite{hkim}, of the turbulent energy transfer.
These phenomenon are consistent with the hairpin vortex model of turbulent structures,
and the large energy transfer may be considered as flowing from the mean shear
into the stretching of the inclined hairpin vortex filaments in the boundary
layer.

The energy imparted to the turbulent development of the fluid, that is, the
growth of hairpin packets, must ultimately arise from the surface via surface
drag. The mean surface stress is \( \overline{\tau _{0}}=\left. \mu \frac{\partial \overline{u_{x}}}{\partial y}\right| _{y=0} \).
The mean velocity profile, \( \overline{u_{x}}(x,y) \), of turbulent boundary
layer flow is strongly affected by the Reynolds stress. The addition of the
Reynolds stress term with \( \overline{u_{x}'u_{y}'}<0 \) in the mean momentum
equation, (\ref{meanmom}), generates a much steeper profile than for laminar
flow, \( \partial _{y}\overline{u_{x}}\gg \partial _{y}u_{x_{laminar}} \),
and hence much greater surface drag.

\subsection{Hairpin vortex contribution to Reynolds stress}

\begin{figure}
{\centering \resizebox*{4.5in}{!}{\includegraphics{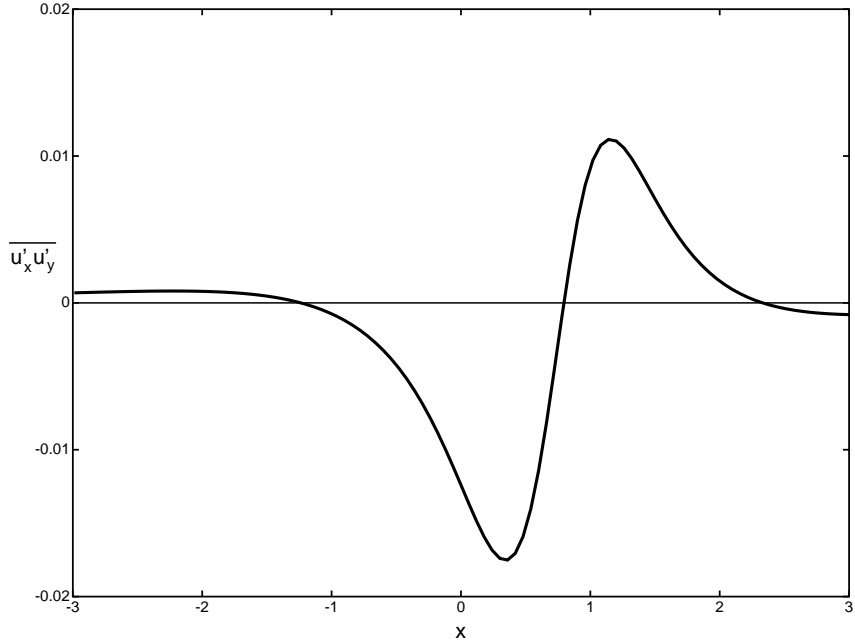}} \par}

\caption{\label{fig:filuv}The Reynolds stress, \protect\( \overline{u_{x}'u_{y}'}\protect \),
as a function of \protect\( x\protect \) along the line \protect\( y=1\protect \),
\protect\( z=0\protect \) threading a sinusoidal filament of unit circulation
centered at \protect\( y=1\protect \), \protect\( x=0\protect \) and oriented
at \protect\( 45^{o}\protect \). The example filament has amplitude \protect\( x_{a}=.75\protect \),
\protect\( y_{a}=.75\protect \) and a wavenumber of \protect\( k=.5\protect \).
The induced velocity of the image filament is also included.}
\end{figure}

It is important to note that the naturally occurring hairpin vortices over a
flat surface, inclined at an approximately \( 45^{o} \) angle, produce the
largest contribution to the Reynolds stress. The legs of such a vortex, oriented
in the \( \pm \left( \widehat{x}+\widehat{y}\right)  \) direction, have large
corresponding fluid velocities in the \( \pm \left( \widehat{x}-\widehat{y}\right)  \)
direction, producing \( \overline{u_{x}'u_{y}'}<0 \). A plot of \( \overline{u_{x}'u_{y}'} \)
along a line threading an example vortex, (Figure \ref{fig:filuv}), shows this
explicitly. The legs of a hairpin vortex, inclined at an an angle \( \theta  \),
will produce a Reynolds stress,
\begin{equation}
\label{rstress}
\overline{u_{x}'u_{y}'}\sim \left( -\sin (\theta )\right) \left( \cos (\theta )\right) =-\frac{1}{2}\sin (2\theta )
\end{equation}
proportional to \( \sin (2\theta ) \).

The velocity field around an advecting hairpin vortex, or vortex packet, directly
explains the observation of streaks, bursts, and sweeps. The streak and burst
events are described by the upwards and backwards flow beneath the vortex filament
heads, while the sweep events are associated with the passing of the vortex
legs near the boundary. Furthermore, the well documented spanwise spacing of
the streaks, \( \lambda \simeq 100wu \) \cite{mhead}, is directly explained
as the wavelength of the maximally unstable mode of the vortex filament instability
that arises from a filament born at \( y_{0}\simeq 6wu \), the top of the high
shear inner region of the boundary layer profile.

A coherent model of turbulent motions over a flat surface is formed by the consideration
of the birth and development of hairpin vortices. The spanwise vortices condense
during the TS instability, and rapidly develop into hairpins, via the three
dimensional instability, to constitute a turbulent boundary layer. Further spanwise
vortices condense in the high shear region, \( y_{0}\simeq 6wu \), via autogeneration,
and evolve, once again via the three dimensional instability, to form hairpin
packets and fill the evolving boundary layer.

\chapter{Filament evolution over a flat boundary\label{sec:flatboundary}}

The dominant structures in the turbulent boundary layer, the hairpin vortices
and hairpin packets, develop via the three dimensional instability of concentrated
patches of spanwise oriented vorticity. The vortex filament model of the developing
turbulent boundary layer provides an elegant framework in which to calculate
the specifics of this instability.

The mean ambient vorticity at the high shear, inner region of the turbulent
boundary layer,
\[
\overrightarrow{\omega ^{a}}\simeq -\widehat{z}\partial _{y}\overline{u_{x}}\simeq -\widehat{z}\frac{u_{*}}{y_{*}}\]
accumulates into the thin, spanwise filaments centered at the filament height,
\( y_{0} \), above the wall. The vorticity of each filament is considered to
condense from a square patch of side \( y_{0} \) centered around the filament
center, producing a vortex with circulation
\[
\Gamma \simeq -\frac{u_{*}}{y_{*}}y_{0}^{2}\]
which propagates in the surrounding, locally constant, mean background flow,
\( u^{b}\simeq \overline{u_{x}}(y_{0})\simeq u_{*}\frac{y_{0}}{y_{*}} \), in
the \( \widehat{x} \) direction.

In the region of the turbulent boundary layer dominated by vortical structures,
above several wall units, the effects of viscosity are small compared to the
dynamical interaction of the vortices with the wall and with themselves. The
simplifying assumption is thus made that the filament propagates in an inviscid
background, \( \mu =0 \), with the no slip condition at the wall relaxed to
the free slip condition, \( u_{y}(y=0)=0 \), or, more generally,
\begin{equation}
\label{freeslip}
\left. \widehat{n}\cdot \overrightarrow{u}\right| _{B}=0
\end{equation}
where \( \widehat{n} \) is the normal at the boundary, \( B \).

\section{The vortex filament}

Consider a single, nearly straight, infinite vortex filament, parallel to the
\( \vec{z} \)-axis, located in space above the flat surface at the position
\begin{equation}
\label{R}
\overrightarrow{R}\left( z,t\right) =\left( X\left( z,t\right) ,Y\left( z,t\right) ,z\right) 
\end{equation}
This time, \( t \), and spanwise coordinate, \( z \), dependent description
of the filament position is adequate for the linear stability analysis in which
the filament is not allowed to loop back on itself in the \( z \) direction. 

Ideally, the filament is represented as an infinitely concentrated region of
vorticity,
\begin{equation}
\label{vortdist}
\overrightarrow{\omega ^{f}}=\Gamma \int ^{\infty }_{-\infty }dz'\, \delta ^{3}(\overrightarrow{x}-\overrightarrow{R}(z',t))\frac{d\overrightarrow{R}}{dz'}
\end{equation}
The actual vorticity distribution is not singular, but is characterized by a
finite vortex core size. This must be considered in calculating the self-interaction
of the filament. However, the above approximation is valid in the case of small
vortex core sizes, and the results of the stability calculation will display
only a weak dependence on core size.

\section{Equations of motion}

The motion of the filament is obtained by using (\ref{vortdist}) in the vorticity
equation, (\ref{vorticity}), and dropping the viscous term to get
\begin{equation}
\label{rmo}
d_{t}\overrightarrow{R}=\partial _{t}\overrightarrow{R}+\left. u_{z}\right| _{\overrightarrow{R}}\, \partial _{z}\overrightarrow{R}=\left. \overrightarrow{u}\right| _{\overrightarrow{R}}
\end{equation}
in which the velocity field, \( \overrightarrow{u}=\overrightarrow{u^{f}}+\overrightarrow{u^{b}} \),
which includes the background flow and the filament induced flow compatible
with the vorticity distribution, is being evaluated at the vortex filament position,
\( \overrightarrow{R}(z,t) \). Inserting (\ref{R}) into (\ref{rmo}) gives
the equations of motion for the \( X(z,t) \) and \( Y(z,t) \) coordinates
of the filament,
\begin{equation}
\label{eomf}
\begin{array}{ccl}
\partial _{t}X & = & -u_{z}\, \partial _{z}X+u_{x}\\
\partial _{t}Y & = & -u_{z}\, \partial _{z}Y+u_{y}
\end{array}
\end{equation}
The flow induced by the vortex filament, \( \overrightarrow{u} \), must be
calculated at each moment from the geometry of the filament and boundary.

\section{Induced flow}

The filament compatible velocity field, \( \overrightarrow{u^{f}} \), must
satisfy \( \overrightarrow{\omega ^{f}}=\overrightarrow{\nabla }\times \overrightarrow{u^{f}} \),
with \( \overrightarrow{\omega ^{f}} \) given by (\ref{vortdist}), as well
as satisfying the free-slip boundary conditions, (\ref{freeslip}). Since the
fluid is considered incompressible, the velocity may be written as the curl
of a stream function, \( \overrightarrow{u}=\overrightarrow{\nabla }\times \overrightarrow{\Psi } \),
with the stream function , \( \overrightarrow{\Psi }=\overrightarrow{\Psi ^{f}}+\overrightarrow{\Psi ^{b}} \),
consisting of terms for the filament induced flow and the background flow, and
background streamfunction \( \overrightarrow{\Psi ^{b}}=u^{b}y\widehat{z} \)
for the current case. This gives, via a vector identity, the relationship between
the vorticity and stream function, 
\[
\overrightarrow{\omega ^{f}}=\overrightarrow{\nabla }\times \overrightarrow{\nabla }\times \overrightarrow{\Psi ^{f}}=-\nabla ^{2}\overrightarrow{\Psi ^{f}}+\overrightarrow{\nabla }\left( \overrightarrow{\nabla }\cdot \overrightarrow{\Psi ^{f}}\right) \]
The gauge choice \( \overrightarrow{\nabla }\cdot \overrightarrow{\Psi ^{f}}=0 \)
may be made without affecting the flow, \( \overrightarrow{u^{f}} \). Finding
the stream function, and hence the induced velocity, is then reduced to the
problem of solving the vector Poisson's equation, \( \nabla ^{2}\overrightarrow{\Psi ^{f}}=-\overrightarrow{\omega ^{f}} \),
with the appropriate boundary conditions. This solution may be written as 
\begin{eqnarray}
\Psi ^{f}_{i} & = & \int d^{3}x'\, G_{ij}(\overrightarrow{x},\overrightarrow{x'})\omega _{j}(\overrightarrow{x'})\nonumber \\
 & = & \Gamma \int dz'\, G_{ij}(\overrightarrow{x},\overrightarrow{R}(z'))\frac{dR_{j}}{dz'}\label{streamfunc} 
\end{eqnarray}
by using the (dyadic) Green function satisfying 
\begin{equation}
\label{poiss}
\nabla ^{2}G_{ij}=-\delta ^{3}(\overrightarrow{x}-\overrightarrow{x'})\delta _{ij}
\end{equation}
and the boundary condition 
\begin{equation}
\label{bcg}
\left. \epsilon _{ijk}n_{j}G_{kl}\right| _{B}=0\quad \forall \left\{ i,l\right\} 
\end{equation}
equivalent to \( \left. \overrightarrow{n}\times \overrightarrow{G}\right| _{B}=0 \).

Green's function satisfying (\ref{poiss}) and (\ref{bcg}) for a flat boundary
at \( y=0 \) is
\begin{equation}
\label{gflat}
\begin{array}{ccc}
G^{F}\left( (x,y,z),(x',y',z')\right)  & = & \frac{1}{4\pi \sqrt{\left( x-x'\right) ^{2}+\left( y-y'\right) ^{2}+\left( z-z'\right) ^{2}}}\left[ \begin{array}{ccc}
1 & 0 & 0\\
0 & 1 & 0\\
0 & 0 & 1
\end{array}\right] \\
 &  & -\frac{1}{4\pi \sqrt{\left( x-x'\right) ^{2}+\left( y+y'\right) ^{2}+\left( z-z'\right) ^{2}}}\left[ \begin{array}{ccc}
1 & 0 & 0\\
0 & -1 & 0\\
0 & 0 & 1
\end{array}\right] 
\end{array}
\end{equation}
The construction of this Green function may be understood via the method of
images. For each vortical element that acts as a vector potential source through
the first term in (\ref{gflat}), there is a mirror image source beneath the
surface that acts through the second term in (\ref{gflat}). The minus sign
in \( G^{F}_{yy} \) arises because the mirrored element points in the opposite
direction in \( \widehat{y} \). The image ensures that the boundary conditions,
(\ref{bcg}), are satisfied, since it cancels out the components of \( \overrightarrow{G^{F}} \)
parallel to the surface.

The velocity field induced by the vortex filament and its image, as well as
the background flow, is found by taking the curl of the stream function to get
\begin{equation}
\label{biosavaart}
\overrightarrow{u}=u^{b}\widehat{x}-\frac{\Gamma }{4\pi }\int dz'\left\{ \frac{\left( \overrightarrow{x}-\overrightarrow{R}(z')\right) }{\left| \overrightarrow{x}-\overrightarrow{R}(z')\right| ^{3}}\times \frac{d\overrightarrow{R}}{dz'}-\frac{\left( \overrightarrow{x}-\overrightarrow{R^{I}}(z')\right) }{\left| \overrightarrow{x}-\overrightarrow{R^{I}}(z')\right| ^{3}}\times \frac{d\overrightarrow{R^{I}}}{dz'}\right\} 
\end{equation}
which is recognizeable as the Bio-Savart law for the vorticity/current source,
with the image vortex position defined as \( \overrightarrow{R^{I}}(z)=\left( X(z),-Y(z),z\right)  \).

\section{Two dimensional motion}

The above equations simplify dramatically for the case of a straight filament.
In this case, the filament position is independent of \( z \), and can be written
as
\[
\overrightarrow{R_{2}}\left( t\right) =\left( x_{0}\left( t\right) ,y_{0}\left( t\right) \right) \]
The equations of motion, (\ref{eomf}), become simply
\[
\begin{array}{ccc}
\partial _{t}x_{0} & = & u_{x}\\
\partial _{t}y_{0} & = & u_{y}
\end{array}\]
and the induced velocity may be found by integrating (\ref{biosavaart}) or
solving Poisson's equation in two dimensions, and differentiating, to get
\[
\begin{array}{ccl}
u_{x} & = & u^{b}-\frac{\Gamma }{2\pi }\frac{\left( y-y_{0}\right) }{\left( x-x_{0}\right) ^{2}+\left( y-y_{0}\right) ^{2}}+\frac{\Gamma }{2\pi }\frac{\left( y+y_{0}\right) }{\left( x-x_{0}\right) ^{2}+\left( y+y_{0}\right) ^{2}}\\
u_{y} & = & \frac{\Gamma }{2\pi }\frac{\left( x-x_{0}\right) }{\left( x-x_{0}\right) ^{2}+\left( y-y_{0}\right) ^{2}}-\frac{\Gamma }{2\pi }\frac{\left( x-x_{0}\right) }{\left( x-x_{0}\right) ^{2}+\left( y+y_{0}\right) ^{2}}
\end{array}\]
Evaluating this induced velocity at the filament in the equations of motion
gives
\begin{equation}
\label{eom2}
\begin{array}{ccl}
\partial _{t}x_{0} & = & u^{b}+\frac{\Gamma }{2\pi }\frac{1}{2y_{0}}\\
\partial _{t}y_{0} & = & 0
\end{array}
\end{equation}
which implies that a straight filament travels over the flat surface at a constant
height, with a constant velocity of \( u^{b}+\frac{\Gamma }{2\pi }\frac{1}{2y_{0}} \)
in the \( \widehat{x} \) direction.

\section{Hamiltonian formulation }

It is interesting and very useful that the dynamics of the vortex filament motion,
(\ref{eomf}), have a Hamiltonian formulation \cite{arouhi}. Taking the dynamical
variables to be \( X(z,t) \) and \( Y(z,t) \) and the canonical, instantaneous,
fundamental Poisson bracket to be
\begin{equation}
\left\{ X(z),Y(z')\right\} =\frac{1}{\Gamma }\delta (z-z')
\end{equation}
the equations of motion are obtained as
\begin{equation}
\label{heom}
\begin{array}{ccccccc}
\partial _{t}X & = & \left\{ X,H\right\}  & = & \frac{1}{\Gamma }\frac{\delta H}{\delta Y} & = & -u_{z}\, \partial _{z}X+u_{x}\\
\partial _{t}Y & = & \left\{ Y,H\right\}  & = & -\frac{1}{\Gamma }\frac{\delta H}{\delta X} & = & -u_{z}\, \partial _{z}Y+u_{y}
\end{array}
\end{equation}
in which the Hamiltonian is the kinetic energy of the filament induced flow
as a functional of \( X \) and \( Y \),
\begin{equation}
\label{ham}
\begin{array}{ccl}
H & = & \frac{1}{2}\int d^{3}x\left| \overrightarrow{u}\right| ^{2}=\frac{1}{2}\int d^{3}x\left( \overrightarrow{u^{f}}\cdot \overrightarrow{u^{f}}+2\overrightarrow{u^{f}}\cdot \overrightarrow{u^{b}}\right) \\
 & = & \frac{1}{2}\int d^{3}x\left( \left| \overrightarrow{\nabla }\times \overrightarrow{\Psi ^{f}}\right| ^{2}+2\left( \overrightarrow{\nabla }\times \overrightarrow{\Psi ^{f}}\right) \cdot \left( \overrightarrow{\nabla }\times \overrightarrow{\Psi ^{b}}\right) \right) \\
 & = & \frac{1}{2}\int d^{3}x\left( \overrightarrow{\omega ^{f}}\cdot \overrightarrow{\Psi ^{f}}+2\overrightarrow{\omega ^{f}}\cdot \overrightarrow{\Psi ^{b}}\right) \\
 & = & \frac{\Gamma ^{2}}{2}\int \int dzdz'\frac{\partial R_{i}}{\partial z}G_{ij}\left( \overrightarrow{R}(z),\, \overrightarrow{R}(z')\right) \frac{\partial R_{j}}{\partial z'}+\Gamma \int dz\frac{d\overrightarrow{R}}{dz}\cdot \overrightarrow{\Psi ^{b}}
\end{array}
\end{equation}
Note the common abuse of notation by which a variational derivative such as
\( \frac{\delta H}{\delta X(z)} \) is defined implicitly such that 
\[
\delta H=\int dz\frac{\delta H}{\delta X(z)}\delta X(z)+\int dz\frac{\delta H}{\delta Y(z)}\delta Y(z)\]

The existence of Hamiltonian dynamics for the motion of the filament will facilitate
the linear stability analysis as well as aid in the calculation of the equations
of motion in new variables via canonical transformations.

\section{Self interaction }

The equations for the three dimensional motion of the filament, (\ref{rmo}),
are presently ill defined because the induced velocity, (\ref{biosavaart}),
diverges at the filament. This is remedied by using a filament with a finite
core size. Rather then make a smoothing correction to the vorticity distribution,
(\ref{vortdist}), with the associated complication of dealing with core dynamics
\cite{rklein}\cite{aqi}, a cutoff is imposed on the self-induction term for
the stream function \cite{psaff}
\[
\Psi ^{f}_{i}=\Gamma \int _{\left[ \delta \right] }dz'\, G_{ij}(\overrightarrow{x},\overrightarrow{R}(z'))\frac{dR_{j}}{dz'}\]
where the \( \left[ \delta \right]  \) stands for the region \( z'<z-a\delta  \)
and \( z'>z+a\delta  \), \( a \) is the vortex core radius, and \( \delta  \)
is a numerical factor. The approximate value of \( \delta  \) is determined
by establishing consistency with the known self induced velocity of a solid
cored vortex ring, and is found to be \( \delta \simeq \frac{1}{2}e^{\frac{1}{4}} \).

For small core sizes this cutoff method is equivalent to the modification of
the Green function to \cite{psaff}
\begin{equation}
\label{sgflat}
\begin{array}{ccc}
G^{F} & = & \frac{1}{4\pi \sqrt{\left( x-x'\right) ^{2}+\left( y-y'\right) ^{2}+\left( z-z'\right) ^{2}+e^{-\frac{3}{2}}a^{2}}}\left[ \begin{array}{ccc}
1 & 0 & 0\\
0 & 1 & 0\\
0 & 0 & 1
\end{array}\right] \\
 &  & -\frac{1}{4\pi \sqrt{\left( x-x'\right) ^{2}+\left( y+y'\right) ^{2}+\left( z-z'\right) ^{2}}}\left[ \begin{array}{ccc}
1 & 0 & 0\\
0 & -1 & 0\\
0 & 0 & 1
\end{array}\right] 
\end{array}
\end{equation}
The use of this approximate, non-singular Green function is superior to the
cutoff method for calculations. It is used in (\ref{ham}) to form the new,
non-singular Hamiltonian.

The radius of the vortex cores, \( a \), for the initial spanwise filaments
must be established by direct experimental observation. An inspection of slides
of turbulent boundary layer flow indicates core sizes on the order of \( a\simeq 1wu \)
\cite{mhead}. For a filament located at an initial height of \( y_{0}\simeq 6wu \)
this corresponds to a relative core radius of
\begin{equation}
\label{coresize}
\frac{a}{y_{0}}=\frac{1}{6}
\end{equation}
Although this quantity is only roughly estimated, the filament instability of
interest, the instability induced by the filament interaction with the wall,
is only weakly dependent on core size.

\section{Linearization\label{sec: linearization}}

The equations of motion may be linearized for small perturbations from the straight
filament solution. An analysis of the filament behavior for small perturbations
will display the nature of the three dimensional instability, including its
angle and growth as a function of spanwise wavelength. This linear analysis
will later be confirmed and extended by a fully nonlinear numerical simulation
in Section \ref{sec:numsim}.

The equations of motion, (\ref{eomf}), are linearized around the two dimensional
filament motion to get,
\begin{equation}
\label{pert}
\begin{array}{ccl}
X\left( z,t\right)  & = & x_{0}\left( t\right) +\varepsilon \widetilde{x}\left( z,t\right) \\
Y\left( z,t\right)  & = & y_{0}(t)+\varepsilon \widetilde{y}\left( z,t\right) 
\end{array}
\end{equation}
where \( \varepsilon  \) is a bookkeeping variable used to keep track of the
order in the small perturbative variables, \( \widetilde{x} \) and \( \widetilde{y} \).
The analysis of linearized motion is facilitated by the Hamiltonian formulation.
The Hamiltonian, (\ref{ham}), for the filament over the flat boundary is expanded
order by order in \( \varepsilon  \) to obtain
\begin{equation}
H^{F}=H^{F\left( 0\right) }+\varepsilon H^{F\left( 1\right) }+\varepsilon ^{2}H^{F\left( 2\right) }+O(\varepsilon ^{3})
\end{equation}
in which \( H^{F\left( 2\right) } \) is the part of the Hamiltonian quadratic
in \( \widetilde{x} \) and \( \widetilde{y} \), and \( H^{F\left( 1\right) } \)
is zero. The equations of motion, (\ref{heom}), are then similarly expanded
order by order in \( \varepsilon  \). The zeroth order equations are the two
dimensional equations of motion, (\ref{eom2}). The first order equations produce
the linearized equations of motion for the filament,
\begin{equation}
\label{leom}
\begin{array}{ccl}
\partial _{t}\widetilde{x} & = & \frac{1}{\Gamma }\frac{\delta H^{F\left( 2\right) }}{\delta \widetilde{y}}\\
\partial _{t}\widetilde{y} & = & -\frac{1}{\Gamma }\frac{\delta H^{F\left( 2\right) }}{\delta \widetilde{x}}
\end{array}
\end{equation}
which depend only on the quadratic part of the Hamiltonian, 
\begin{equation}
\label{qham}
H^{F\left( 2\right) }=\frac{\Gamma ^{2}}{2}\int \int dzdz'\left( \frac{\partial \widetilde{R_{\alpha }}}{\partial z}G_{\alpha \beta }^{F\left( 0\right) }\frac{\partial \widetilde{R_{\beta }}}{\partial z'}+G_{zz}^{F\left( 2\right) }\right) 
\end{equation}
in which \( \widetilde{\overrightarrow{R}}=\left( \widetilde{x},\widetilde{y}\right)  \),
Greek indices \( \left\{ \alpha ,\beta \right\}  \) range only over \( \left\{ x,y\right\}  \),
and \( G^{F\left( 0\right) }_{\alpha \beta } \) and \( G^{F\left( 2\right) }_{zz} \)
are the modified Green function, (\ref{sgflat}), components of zeroth and quadratic
order in \( \varepsilon  \),
\[
\begin{array}{ccl}
G^{F\left( 0\right) }_{\alpha \beta } & = & \left. G^{F}_{\alpha \beta }(\overrightarrow{R}(z),\overrightarrow{R}(z'))\right| _{\varepsilon =0}\\
G^{F\left( 0\right) }_{zz} & = & \left. \frac{1}{2}\frac{\partial ^{2}}{\partial \varepsilon ^{2}}G^{F}_{zz}(\overrightarrow{R}(z),\overrightarrow{R}(z'))\right| _{\varepsilon =0}
\end{array}\]

\section{Diagonalization\label{sec: diag}}

The linearized equations of motion, (\ref{leom}), along with the quadratic
Hamiltonian, (\ref{qham}), comprise a set of linear, integro-partial differential
equations for the dynamical variables \( \widetilde{x}(z,t) \) and \( \widetilde{y}(z,t) \).
This set of equations is reduced to a set of linear, ordinary differential equations
by applying a Fourier transform in \( z \) to obtain the new complex dynamical
variables, \( \widetilde{x}\left( k,t\right)  \), defined as
\[
\widetilde{x}(k,t)=\frac{1}{2\pi }\int ^{\infty }_{-\infty }dz\, e^{-ikz}\widetilde{x}(z,t)\]
with
\[
\widetilde{x}(z,t)=\int ^{\infty }_{-\infty }dk\, e^{ikz}\widetilde{x}(k,t)\]
and \( \widetilde{y}(k,t) \) defined similarly. This is a canonical transformation.
The new fundamental Poisson bracket becomes
\begin{equation}
\label{fpb}
\begin{array}{ccl}
\left\{ \widetilde{x}(k),\widetilde{y}^{*}(k')\right\}  & = & \frac{1}{\Gamma }\int dz\left\{ \frac{\delta \widetilde{x}(k)}{\delta \widetilde{x}(z)}\frac{\delta \widetilde{y}^{*}(k')}{\delta \widetilde{y}(z)}-\frac{\delta \widetilde{y}^{*}(k')}{\delta \widetilde{x}(z)}\frac{\delta \widetilde{x}(k)}{\delta \widetilde{y}(z)}\right\} \\
 & = & \frac{1}{2\pi \Gamma }\delta (k-k')
\end{array}
\end{equation}

The next step is to show how the quadratic Hamiltonian, (\ref{qham}), is diagonalized
by this transformation, and calculate the resulting equations of motion. Since
the Green function, (\ref{sgflat}), within the Hamiltonian is explicitly dependent
only on \( z-z' \) and not on \( z+z' \), it is useful to write it as a function
of \( \xi \equiv z-z' \) and perform the Fourier transform in this variable
to obtain
\[
G^{F}(x,y,\xi ,x',y')=\int dk\, e^{ik\xi }G^{F}(x,y,k,x',y')\]
in which the Fourier transformed Green function is calculated as
\[
\begin{array}{ccl}
G^{F}(x,y,k,x',y') & = & \frac{1}{2\pi }\int d\xi \, e^{-ik\xi }G^{F}(x,y,\xi ,x',y')\\
 & = & \frac{1}{\left( 2\pi \right) ^{2}}K_{0}(k\sqrt{\left( x-x'\right) ^{2}+\left( y-y'\right) ^{2}+e^{-\frac{3}{2}}a^{2}})\left[ \begin{array}{ccc}
1 & 0 & 0\\
0 & 1 & 0\\
0 & 0 & 1
\end{array}\right] \\
 &  & -\frac{1}{\left( 2\pi \right) ^{2}}K_{0}(k\sqrt{\left( x-x'\right) ^{2}+\left( y+y'\right) ^{2}})\left[ \begin{array}{ccc}
1 & 0 & 0\\
0 & -1 & 0\\
0 & 0 & 1
\end{array}\right] 
\end{array}\]
with \( K_{0} \) the modified Bessel function. The quadratic Hamiltonian is
then written as
\[
\begin{array}{ccl}
H^{F\left( 2\right) } & = & \frac{\Gamma ^{2}}{2}\int \int \int dzdz'dk\, e^{ik\xi }\left\{ \frac{\partial \widetilde{R_{\alpha }}}{\partial z}G_{\alpha \beta }^{F\left( 0\right) }(X(z),Y(z),k,X(z'),Y(z'))\frac{\partial \widetilde{R_{\beta }}}{\partial z'}\right. \\
 &  & \left. +G_{zz}^{F\left( 2\right) }(X(z),Y(z),k,X(z'),Y(z'))\right\} 
\end{array}\]
The \( G^{F\left( 0\right) } \) and \( G^{F\left( 2\right) } \) are computed,
the Fourier integrals are substituted for \( \widetilde{x}(z) \) and \( \widetilde{y}(z) \),
and the resulting quintiple integral vanishes in a poof of delta functions to
give the diagonalized Hamiltonian in the Fourier variables,
\begin{equation}
\label{fham}
H^{F\left( 2\right) }=\frac{\Gamma ^{2}}{2}\int dk\, k^{2}\left\{ A(k)\, \widetilde{x}^{*}(k,t)\, \widetilde{x}\left( k,t\right) +B(k)\, \widetilde{y}^{*}(k,t)\, \widetilde{y}(k,t)\right\} 
\end{equation}
with
\begin{equation}
\label{AB}
\begin{array}{ccl}
A & = & S(ka)+\left\{ \frac{1}{\left( 2ky_{0}\right) ^{2}}-K_{0}(2ky_{0})-\frac{1}{2ky_{0}}K_{1}(2ky_{0})\right\} \\
B & = & S(ka)+\left\{ -\frac{1}{\left( 2ky_{0}\right) ^{2}}-\frac{1}{2ky_{0}}K_{1}(2ky_{0})\right\} 
\end{array}
\end{equation}
and the vortex filament self interaction term
\begin{equation}
\label{Ska}
\begin{array}{ccl}
S(ka) & = & -\frac{1}{\left( e^{-\frac{3}{4}}ka\right) ^{2}}+K_{0}(e^{-\frac{3}{4}}ka)+\frac{1}{e^{-\frac{3}{4}}ka}K_{1}(e^{-\frac{3}{4}}ka)\\
 & \simeq  & 0.183-\frac{1}{2}\ln (ka)+O(\left( ka\right) ^{2})
\end{array}
\end{equation}
The linear equations of motion for each mode are then easily computed from (\ref{fham})
and (\ref{fpb}) to be
\begin{equation}
\label{lfeom}
\begin{array}{ccccc}
\partial _{t}\widetilde{x}(k,t) & = & \frac{1}{2\pi \Gamma }\frac{\delta H^{F\left( 2\right) }}{\delta \widetilde{y}^{*}} & = & \frac{\Gamma k^{2}}{2\pi }B(k)\, \widetilde{y}\\
\partial _{t}\widetilde{y}(k,t) & = & -\frac{1}{2\pi \Gamma }\frac{\delta H^{F\left( 2\right) }}{\delta \widetilde{x}^{*}} & = & -\frac{\Gamma k^{2}}{2\pi }A(k)\, \widetilde{x}
\end{array}
\end{equation}

\section{Linear instability}

The set of linear ordinary differential equations of motion for each mode, (\ref{lfeom}),
are reduced to an eigenvalue problem by assuming a solution of the form
\[
\begin{array}{ccc}
\widetilde{x} & = & \widetilde{x_{1}}e^{\sigma t}\\
\widetilde{y} & = & \widetilde{y_{1}}e^{\sigma t}
\end{array}\]
to get
\[
\sigma \left[ \begin{array}{c}
\widetilde{x_{1}}\\
\widetilde{y_{1}}
\end{array}\right] =\left[ \begin{array}{cc}
0 & \frac{\Gamma k^{2}}{2\pi }B(k)\\
-\frac{\Gamma k^{2}}{2\pi }A(k) & 0
\end{array}\right] \left[ \begin{array}{c}
\widetilde{x_{1}}\\
\widetilde{y_{1}}
\end{array}\right] \]
The two resulting eigenvalues are
\begin{equation}
\label{eigenval}
\sigma _{\pm }(k)=\pm \left| \frac{\Gamma k^{2}}{2\pi }\right| \sqrt{-AB}
\end{equation}
with the angle, up from \( \widehat{x} \), of the corresponding eigenvectors
equal to
\[
\theta _{\pm }=\arctan (\frac{\widetilde{y_{1}}}{\widetilde{x_{1}}})=\arctan (\pm \frac{\Gamma }{\left| \Gamma \right| }\frac{\sqrt{-AB}}{B})\]

\begin{figure}
{\centering \resizebox*{4.5in}{!}{\includegraphics{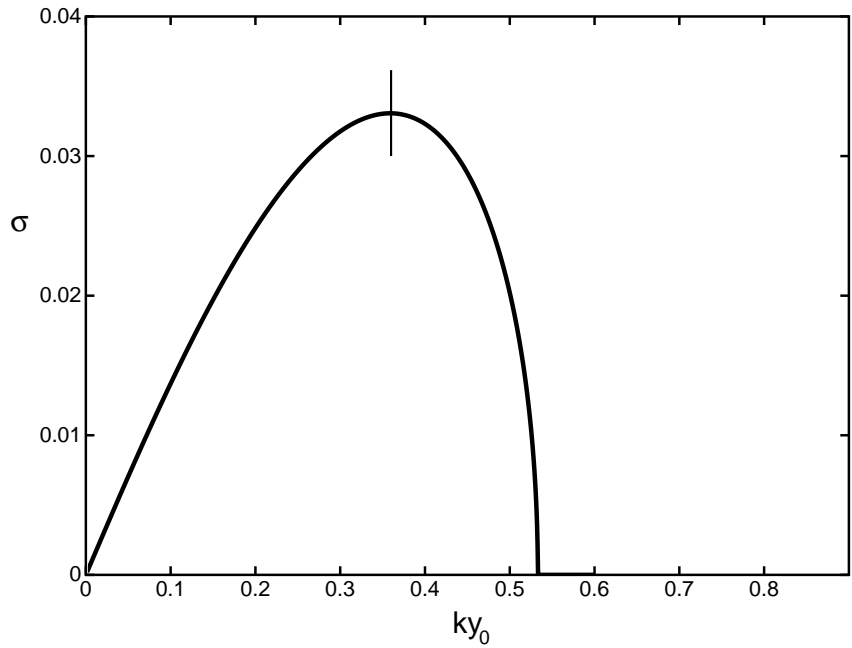}} \par}

\caption{\label{fig:stab}The growth parameter, \protect\( \sigma _{+}\protect \),
in units of \protect\( \frac{\left| \Gamma \right| }{y_{0}^{2}}\protect \),
as a function of spanwise wavenumber, \protect\( ky_{0}\protect \). The maximum
at \protect\( ky_{0}\simeq 0.36\protect \) is marked. The core size was set
to be \protect\( \frac{a}{y_{0}}=\frac{1}{6}\protect \) to obtain this plot.}
\end{figure}

\begin{figure}
{\centering \resizebox*{4.5in}{!}{\includegraphics{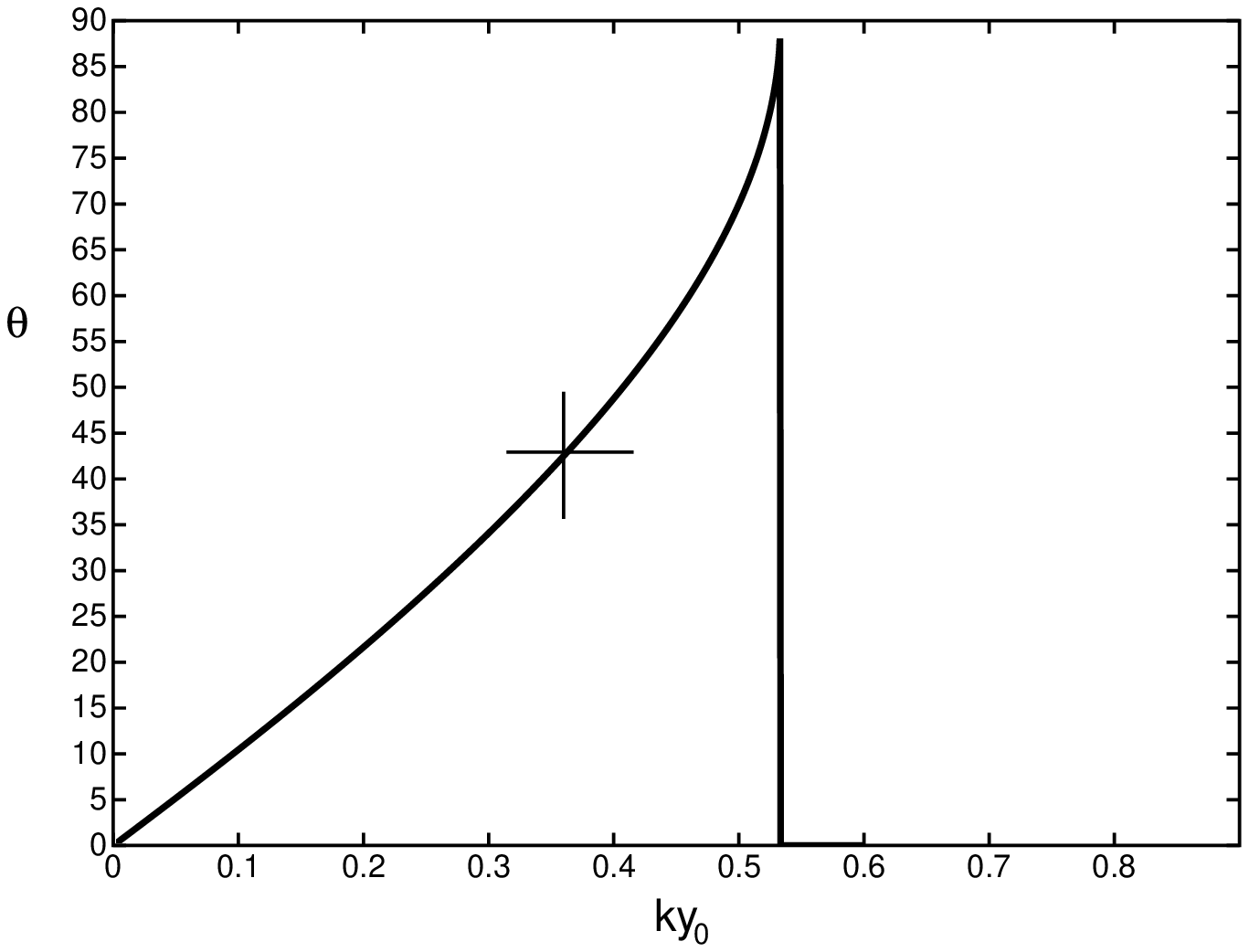}} \par}

\caption{\label{fig:ang}The instability angle, \protect\( \theta _{+}\protect \),
as a function of spanwise wavenumber, \protect\( ky_{0}\protect \). The maximally
unstable wavenumber, \protect\( ky_{0}\simeq 0.36\protect \), is marked, indicating
the angle of the maximally unstable mode to be \protect\( \theta _{+}\simeq 43^{o}\protect \).
The core size was set to be \protect\( \frac{a}{y_{0}}=\frac{1}{6}\protect \)
to obtain this plot.}
\end{figure}

The positive eigenvalue, or growth parameter, \( \sigma _{+} \), is plotted
in Figure \ref{fig:stab}. All small wavenumber perturbations of the filament are unstable up to \( ky_{0}\simeq .53 \),
with the maximum instability occurring at \( ky_{0}\simeq .36 \). For wavenumbers
above \( .53 \) the eigenvalues, (\ref{eigenval}), become imaginary, corresponding
to filament perturbations that rotate rather then grow. The unstable eigenmodes
corresponding to \( \sigma _{+} \) grow at an angle, \( \theta _{+} \), plotted
in Figure \ref{fig:ang}. All unstable modes grow in planes with angles of \( 0^{o} \) to \( 90^{o} \)
up from \( \widehat{x} \), with the maximally unstable mode growing at an angle
of \( \theta _{+}\simeq 43^{o} \). The wavenumber of the maximally unstable
mode is dependent on the choice of core size, (\ref{coresize}), for the calculation.
However, the instability is only weakly dependent on this choice. For example,
a core of only one tenth the current estimated size of \( \frac{a}{y_{0}}=\frac{1}{6} \)
produces a similar instability plot with the maximum at \( ky_{0}\simeq 0.28 \).
Also, the angle of the maximally unstable mode remains \( \theta _{+}\simeq 43^{o} \)
over several orders of magnitude of core size.

For any small initial perturbation of the straight filament, the high wavenumber
modes rotate about the core while the low wavenumber modes expand in the unstable
plane, \( \theta _{+} \), and contract in the stable plane, \( \theta _{-} \).
The maximally unstable mode, \( ky_{0}\simeq 0.36 \), expands exponentially
faster then the others, and rapidly becomes the dominant deformation of the
filament. The filament evolves into a sinusoidal shape inclined at approximately
\( 43^{o} \), with a wavelength of \( \lambda =\frac{2\pi }{k}\simeq 17y_{0} \),
as plotted in Figure \ref{fig:filament}. Although the linear analysis gives
a precise description of vortex filament motion for small perturbations, the
full equations of motion, (\ref{eomf}), are highly nonlinear; and it is expected
that nonlinear mechanisms will determine the dynamics after the initial perturbations
grow to significant size.

\section{Numerical simulation and nonlinear evolution\label{sec:numsim} }

Numerical simulation provides an effective means of exploring nonlinear behavior,
as well as confirming linear analysis. The fully nonlinear equations of motion
for a vortex filament lend themselves readily to numerical analysis. However,
the equation for vortex position, (\ref{R}), must be modified to a fully parametric
representation to allow for the possibility of the filament doubling back in
the \( \widehat{z} \) direction,
\[
\overrightarrow{R}(s,t)=\left( X(s,t),Y(s,t),Z(s,t)\right) \]
in which \( s \) is an arbitrary parameter along the filament. The equation
of motion for the filament is then simply
\[
\partial _{t}\overrightarrow{R}(s,t)=\left. \overrightarrow{u}\right| _{\overrightarrow{R}}\]
with the velocity field given by the Bio-Savart integral, (\ref{biosavaart}),
with a finite core size. The full set of nonlinear integro-partial differential
equations is
\[
\begin{array}{ccc}
\partial _{t}\overrightarrow{R}(s,t) & = & u^{b}\widehat{x}-\frac{\Gamma }{4\pi }\int ds'\left\{ \frac{\left( \overrightarrow{R}(s,t)-\overrightarrow{R}(s',t)\right) }{\left( \left| \overrightarrow{R}(s,t)-\overrightarrow{R}(s',t)\right| ^{2}+e^{-\frac{3}{2}}a^{2}\right) ^{\frac{3}{2}}}\times \partial _{s'}\overrightarrow{R}(s',t)\right. \\
 &  & \left. -\frac{\left( \overrightarrow{R}(s,t)-\overrightarrow{R^{I}}(s',t)\right) }{\left| \overrightarrow{R}(s,t)-\overrightarrow{R^{I}}(s',t)\right| ^{3}}\times \partial _{s'}\overrightarrow{R^{I}}(s',t)\right\} 
\end{array}\]
with the image position \( \overrightarrow{R^{I}}(s,t)=\left( X(s,t),-Y(s,t),Z(s,t)\right)  \).
These equations are then discretized in \( s \) and integrated forward in time.
Care is taken to insert new discretization points along the filament as it is
stretched, to preserve the accuracy of the scheme.

\begin{figure}
{\centering \resizebox*{5.5in}{!}{\includegraphics{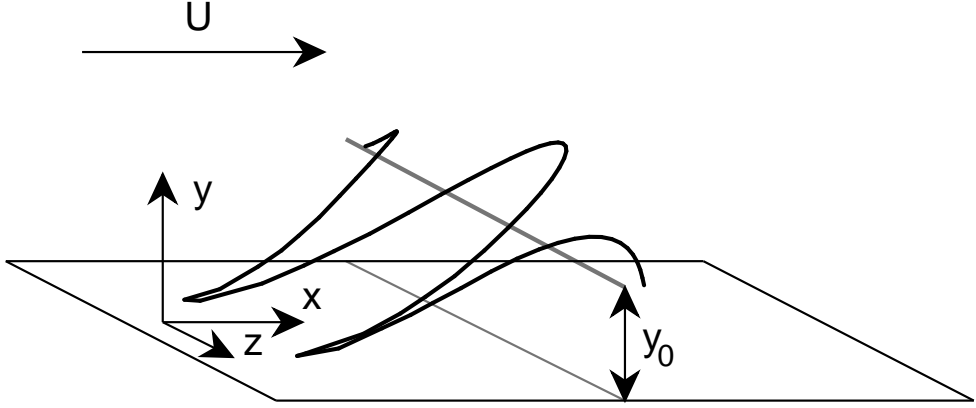}} \par}

\caption{\label{fig:filamentnl}The vortex filament evolving nonlinearly from a small
initial Gaussian perturbation. This figure may be compared directly with Figure
\ref{fig:filament}.}
\end{figure}

An example of the resulting filament motion is shown in Figure \ref{fig:filamentnl}. For this case, the initial filament was taken to have a small Gaussian perturbation
in the \( \widehat{x} \) direction, \( \overrightarrow{R}(s,0)=\left( .1e^{-\left( \frac{s}{4}\right) ^{2}},1,s\right)  \),
to simulate the introduction of a small disturbance to the flow. This Gaussian
has a small projection onto the maximally unstable mode of wavelength \( \lambda \simeq 17y_{0} \),
which rapidly grows to dominate the filament geometry, confirming the accuracy
of the linear analysis. The nonlinear effects begin to contribute significantly
to the evolution as the legs of the filament approach the surface. The legs
are drawn out in the \( -\widehat{x} \) direction, producing counter rotating
streamwise vortices near the wall. And the initially planar sinusoidal curve
of the filament becomes slightly bowed as it expands.

Although this numerical simulation gives a satisfactory description of filament
motion for short times, it is insufficient to address long time evolution questions
such as the details of filament autogeneration and multiple filament interaction.
A simulation of autogeneration along these lines would require more detailed
core dynamics as well as the use of vortex filament surgery. Alternatively,
a three dimensional simulation of the Eulerian vorticity field of an autogenerating
hairpin produces excellent results \cite{jzhou}. However, the present numerical
model serves as a successful confirmation of the linear result, displaying the
dominant evolution of \( \theta _{+}\simeq 43^{o} \) modes with wavelength
\( \lambda \simeq 17y_{0} \), and provides a consistent picture of the evolution
of single hairpin vortex features.

\section{Comparison with coherent structures in the turbulent boundary layer}

The qualitative agreement between the analytic results and the reported experimental
observations of hairpin vortices leaves little doubt that this filament instability
is responsible for the development of hairpin vortices in the turbulent boundary
layer. The observed \( 45^{o} \) inclination is given a solid mathematical
foundation, and the numerical simulation of the evolving filament, \ref{fig:filamentnl},
produces a geometry that conforms excellently with the reports of observers
and conceptual models as reviewed by Robinson \cite{srobins}, and more recently
elucidated by Zhong and others \cite{jzhong}, and by Delo and Smits \cite{cdelo}.

The analytic calculation of the existence of the maximally unstable wavenumber,
\( ky_{0}\simeq .36 \), together with the experimental observation of spanwise
wavelengths of \( 100wu \) for these structures in the boundary layer, demands
that the average initial height of the vortex filaments be equal to \( y_{0}\simeq 6wu. \)
This is reasonably located in a region of high mean shear in the turbulent boundary
layer, and agrees well with the measured location of maximal turbulent energy
production \cite{jbalint}.

However, the true power of the newly established mathematical model for vortex
filament dynamics in the boundary layer lies in its potential for describing
and predicting new phenomenon. In the next chapter the existant mathematical
framework will be extended to the case of a corrugated boundary surface, and
new predictions will be extracted regarding the subsequent dynamical behavior
of the vortex filament.

\chapter{Filament evolution over a wavy boundary\label{sec: wavy}}

\begin{quotation}
Several decades of intense academic study have provided an extensive knowledge
base of the simple canonical case. The immediate need is to learn to utilize
this store of information in the context of boundary-layer modeling and control
methodologies, with the eventual goal of practical application to engineering
problems involving real-world, noncanonical boundary layers. -- Stephen K. Robinson
\cite{srobins}
\end{quotation}
In the last paragraph of his excellent review article on coherent motions in
the turbulent boundary layer, Stephen K. Robinson urged researchers to build
upon the foundation of knowledge in vortex filament behavior and to forge ahead
into the uncharted territory of control methodologies. This chapter addresses
one such methodology: the use of a static, wavy boundary to control vortex filament
evolution.

The study of this control methodology is motivated by the existence of such
surface corrugations on the bodies of dolphins, with the physical parameters
of background flow speed and surface ridge geometry taken directly from measurements
of these animals. A study of vortex filament dynamics above such a boundary,
using the tools developed in Chapter \ref{sec:flatboundary}, elucidates the
effects on turbulent boundary layer development and the potential for drag reduction.

A vortex filament propagating over a wavy surface is subjected to rapid periodic
forcing due to its interaction with the surface. The filament is near the surface
over the ridge peaks and far from the surface over the ridge troughs. Also,
the unstable manifold, shown in Figure \ref{fig:velocity}, oscillates in direction
and magnitude as the surface beneath the filament is inclined to the horizontal.
This rapid parametric forcing is similar to the pondermotive forcing of a charged
particle in an electromagnetic wave, and dynamically similar to the parametric
forcing of an inverted pendulum \cite{mlevi},\cite{jblack}. The pondermotive
forcing has a dramatic effect on the instability, as it does in the case of
the pendulum.

\section{The boundary and the vortex filament}

The wavy boundary is taken to be an infinite surface, \( \overrightarrow{B} \),
beneath the filament, uniform in the \( \vec{z} \) direction, and periodic
in the \( \widehat{x} \) direction. Hence,
\[
\overrightarrow{B}(x,z)=\left( x,B(x),z\right) \]
with \( B(x) \) a smooth, periodic function with period \( L \) and trough
to peak ridge height \( h \). The surface normal is
\[
\widehat{n}=\frac{1}{\sqrt{1+B'^{2}}}\left( -B'(x)\widehat{x}+\widehat{y}\right) \]
and the free slip boundary condition, \( \left. \widehat{n}\cdot \overrightarrow{u}\right| _{\overrightarrow{B}}=0 \),
is assumed.

The cutaneous ridges of \textit{Tursiops Truncatus}, (Figure \ref{fig: skin}),
are near sinusoidal, with slightly steeper ridges then troughs. The ridges of
a live dolphin at rest are measured, \( 25cm \) behind the blowhole \cite{pshoe},
to have an average height of \( h\simeq .015mm \) and an average spacing of
\( L\simeq .61mm \), producing a height to length ratio of \( \frac{h}{L}\simeq .025 \).
It is expected that a dolphin will increase or decrease the ridge height, using
cutaneous muscle, to accommodate changes in swimming speed, with the ridge height
to length ratio possibly ranging up to \( \frac{h}{L}\simeq 0.2 \).

Dolphins swim at a cruising speed of \( 10\frac{km}{hr}\simeq 3\frac{m}{s} \),
with a top speed of \( 50\frac{km}{hr}\simeq 14\frac{m}{s} \). For a dolphin
swimming at \( U=3\frac{m}{s} \), the friction velocity at, and slightly beyond,
the transition point is \( u_{*}\simeq \frac{1}{27}U\simeq .11\frac{m}{s} \).
A wall unit at this point is thus \( y_{*}=\frac{\nu }{u_{*}}\simeq .009mm \),
and the vortex filaments are assumed to form at a height of \( y_{0}\simeq 6wu\simeq .054mm \).
The vortices at this height propagate in a background flow field of velocity
\[
u^{b}\simeq u_{*}\frac{y_{0}}{y_{*}}\simeq .66\frac{m}{s}\]
and have a circulation of
\[
\Gamma \simeq -\frac{u_{*}}{y_{*}}y_{0}^{2}=-\nu \left( \frac{y_{0}}{y_{*}}\right) ^{2}\simeq -3.6\times 10^{-5}\frac{m^{2}}{s}\]
Although these initial values for the filaments are calculated for a flat plate,
they will be taken as the approximate values for filaments evolving over a wavy
boundary.

The filament position as a function of \( z \) and \( t \) is, for the case
of a wavy boundary, still written as 
\begin{equation}
\label{R2}
\overrightarrow{R}\left( z,t\right) =\left( X\left( z,t\right) ,Y\left( z,t\right) ,z\right) 
\end{equation}
with the dynamical variables \( X(z,t) \) and \( Y(z,t) \) satisfying the
equations of motion, (\ref{eomf}),
\begin{equation}
\label{eomf2}
\begin{array}{ccc}
\partial _{t}X & = & -u_{z}\, \partial _{z}X+u_{x}\\
\partial _{t}Y & = & -u_{z}\, \partial _{z}Y+u_{y}
\end{array}
\end{equation}
However, the induced velocity, as well as the background velocity field, differs
from the field for a flat plate, (\ref{biosavaart}), because of the wavy boundary.

\section{Orthogonal curvilinear coordinates}

The existence of a wavy boundary significantly complicates the calculation of
the induced velocity field over the surface. However, the problem is simplified
by the introduction of orthogonal curvilinear coordinates that match the boundary.

\begin{figure}
{\centering \resizebox*{5.5in}{!}{\includegraphics{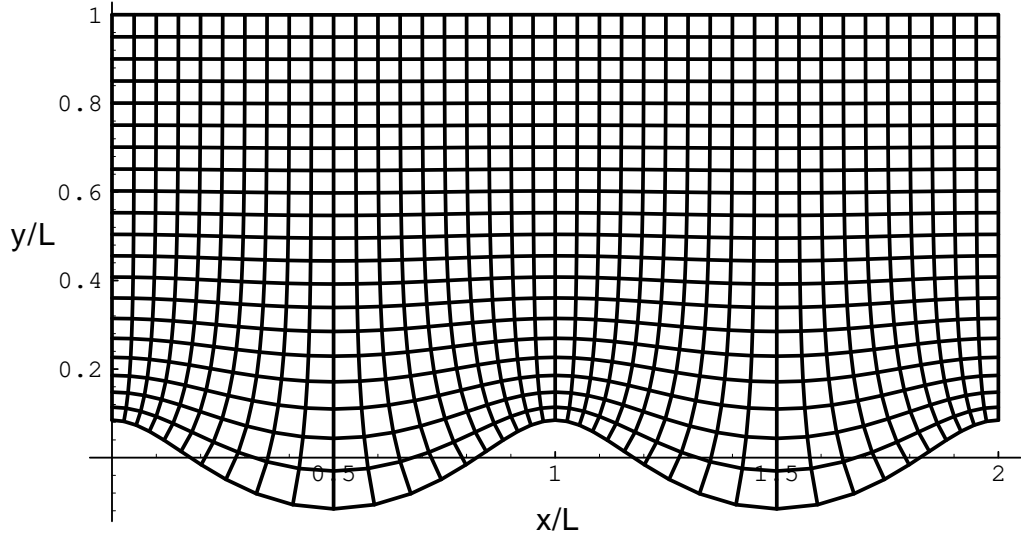}} \par}

\caption{\label{fig: conform} The conformal coordinate mapping of lines of constant
\protect\( u\protect \) and \protect\( v\protect \) into the \protect\( xy\protect \)
plane, with a chosen height to length ratio of \protect\( \frac{h}{L}=.2\protect \).}
\end{figure}

A coordinate transformation is made from \( x \) and \( y \) to new coordinates,
\( u \) and \( v \), such that
\begin{equation}
\label{coord}
\begin{array}{rcl}
x(u,v) & = & \frac{L}{\pi }\arctan \left( \tan \left( \frac{\pi u}{L}\right) \tanh \left( \frac{\pi }{L}\left( v+v_{B}\right) \right) \right) +2Lp\\
y(u,v) & = & \frac{L}{2\pi }\ln \left( 4\cos ^{2}\left( \frac{\pi u}{L}\right) \cosh ^{2}\left( \frac{\pi }{L}\left( v+v_{B}\right) \right) \right. \\
 &  & \left. +4\sin ^{2}\left( \frac{\pi u}{L}\right) \sinh ^{2}\left( \frac{\pi }{L}\left( v+v_{B}\right) \right) \right) -v_{B}
\end{array}
\end{equation}
in which the term with \( p \), equal to the integer part of \( \frac{1}{2}\left( \frac{u}{L}+1\right)  \),
is necessary to give a smoothly increasing \( x \) for \( u>L \), and \( v_{B} \)
is the ridge height parameter, related to the ridge height, \( h \), by
\[
v_{B}=-\frac{L}{2\pi }\ln \left( \tanh (\frac{\pi h}{2L})\right) \]
The coordinate lines of constant \( u \) and \( v \), mapped into the \( xy \)
plane, are shown in Figure \ref{fig: conform}. This coordinate transformation
approximates the wavy boundary of dolphin skin,
\[
\left. B\left( x(u,v)\right) \right| _{v=0}=\left. y(u,v)\right| _{v=0}\]
with peaks slightly sharper then troughs.

The transformation, (\ref{coord}), is a conformal transformation derived from
the complex analytic mapping,
\begin{equation}
\label{zeta}
\zeta (w=u+iv)=x+iy=\frac{iL}{\pi }\ln \left( 2\cos \left( \frac{\pi }{L}\left( u+iv+iv_{B}\right) \right) \right) +2Lp-iv_{B}
\end{equation}
and thus preserves the orthogonality of the coordinates. The great advantage
of using a conformal transformation is the ease of calculating the two dimensional
Laplace operator, and solutions to the Laplace equation, in the new coordinates.
Because the analytic function, (\ref{zeta}), satisfies the Cauchy-Riemann equations,
\[
\begin{array}{ccc}
\partial _{u}x & = & \partial _{v}y\\
\partial _{v}x & = & -\partial _{u}y
\end{array}\]
the two dimensional Laplace operator, in the new coordinates, becomes
\[
\nabla ^{2}=\partial _{x}^{2}+\partial _{y}^{2}=\frac{1}{s^{2}(u,v)}\left( \partial _{u}^{2}+\partial _{v}^{2}\right) \]
in which the function \( s^{2}(u,v) \) is the scaling factor, or Jacobian,
of the conformal transformation -- a measure of the area change of the new curvilinear
coordinate grid -- and is equal to
\[
\begin{array}{ccl}
s^{2}(u,v) & = & \left| \zeta '(w)\right| ^{2}=\left( \partial _{u}x\right) \left( \partial _{v}y\right) -\left( \partial _{u}y\right) \left( \partial _{v}x\right) \\
 & = & \frac{2\cosh \left( \frac{2\pi }{L}\left( v+v_{B}\right) \right) }{\cosh \left( \frac{2\pi }{L}\left( v+v_{B}\right) \right) +\cos \left( \frac{2\pi u}{L}\right) }-1
\end{array}\]
for the particular coordinate change, (\ref{coord}).

Using the new coordinates, the background potential flow satisfying the free-slip
boundary conditions over the wavy surface may be obtained from the simple vector
potential, \( \overrightarrow{\Psi ^{b}}(u,v)=u^{b}v\widehat{z} \). This vector
potential produces an irrotational background velocity field, \( \overrightarrow{u^{b}}=\overrightarrow{\nabla }\times \overrightarrow{\Psi ^{b}} \),
which satisfies the incompressibility condition, \( \overrightarrow{\nabla }\cdot \overrightarrow{u^{b}}=0 \),
as well as the free-slip condition, \( \left. \widehat{n}\cdot \overrightarrow{u^{b}}\right| _{\overrightarrow{B}}=0 \).
The lines of constant \( v \) in the new coordinate system correspond to streamlines
of this background flow over the wavy surface.

Although it is possible to continue to consider the motion of the filament in
cartesian coordinates above the wavy boundary, using the curvilinear coordinates
only for the purpose of velocity field computation, the analysis is greatly
simplified by transforming the dynamical variables of the filament, \( X \)
and \( Y \), into the new coordinate system as well. This obviates the need
to convert back and forth between curvilinear and cartesian coordinates.

\section{Non-canonical transformation}

A non-canonical transformation is made from the old dynamical filament variables,
\( X \) and \( Y \), to the new variables, \( U(X,Y) \) and \( V(X,Y) \),
via the inverse of the coordinate transformation, (\ref{coord}). The fundamental
Poisson bracket in the new coordinates becomes
\begin{equation}
\label{pbw}
\begin{array}{ccl}
\left\{ U(z),V(z')\right\}  & = & \int \int dz''dz'''\left\{ \frac{\delta U(z)}{\delta X(z'')}\frac{\delta V(z')}{\delta Y(z''')}-\frac{\delta V(z')}{\delta X(z'')}\frac{\delta U(z)}{\delta Y(z''')}\right\} \frac{1}{\Gamma }\delta (z''-z''')\\
 & = & \frac{1}{\Gamma s^{2}(U,V)}\delta (z-z')
\end{array}
\end{equation}
with the scale factor, \( s^{2} \), appearing as a direct result of the compression
and expansion of fluid as it is carried over the wavy surface. The position
of the vortex filament in \( uvz \) coordinates is now given by
\begin{equation}
\label{Rw}
\overrightarrow{R}\left( z,t\right) =\left( U\left( z,t\right) ,V\left( z,t\right) ,z\right) _{uvz}
\end{equation}
and the Hamiltonian, (\ref{ham}), in the new coordinates is written as
\begin{equation}
\label{hamw}
H^{W}=\frac{\Gamma ^{2}}{2}\int \int dzdz'\frac{\partial R_{i}}{\partial z}G^{W}_{ij}\left( \overrightarrow{R}(z),\, \overrightarrow{R}(z')\right) \frac{\partial R_{j}}{\partial z'}+\Gamma \int dz\frac{dR_{i}}{dz}\Psi _{i}^{b}
\end{equation}
with the indices now ranging over \( \left\{ u,v,z\right\}  \), and \( G^{W} \)
the new Green function for the wavy boundary.

The boundary conditions for the wavy boundary Green function,
\begin{equation}
\label{bcgw}
\left. \epsilon _{ijk}n_{j}G^{W}_{kl}\right| _{B}=0\quad \forall \left\{ i,l\right\} 
\end{equation}
take a particularly simple form in the curvilinear coordinates, in which \( \widehat{n}=\widehat{v} \),
with (\ref{bcgw}) simplifying to
\begin{equation}
\label{bcgws}
\left. G_{ul}^{W}\right| _{v=0}=\left. G_{zl}^{W}\right| _{v=0}=0\quad \forall \left\{ l\right\} 
\end{equation}
which is \( \left. \widehat{v}\times \overrightarrow{G^{W}}\right| _{v=0}=0 \).
The wavy Green function must also satisfy Poisson's equation,
\begin{equation}
\label{poissw}
\nabla _{ik}^{2}G^{W}_{kj}=-\delta ^{3}(\overrightarrow{x}-\overrightarrow{x'})\delta _{ij}
\end{equation}
with the three dimensional Laplace operator in curvilinear coordinates taking
the form
\begin{equation}
\label{lapw}
\begin{array}{ccl}
\nabla ^{2} & = & \left( \overrightarrow{\nabla }\cdot \right) -\overrightarrow{\nabla }\times \overrightarrow{\nabla }\times \\
 & = & \frac{1}{s^{2}}\left( \partial ^{2}_{u}+\partial ^{2}_{v}+s^{2}\partial ^{2}_{z}\right) \left[ \begin{array}{ccc}
1 & 0 & 0\\
0 & 1 & 0\\
0 & 0 & 1
\end{array}\right] \\
 &  & +\frac{1}{s^{3}}\left( \left( \partial ^{2}_{u}s\right) -\frac{2}{s}\left( \partial _{u}s\right) ^{2}+\left( \partial ^{2}_{v}s\right) -\frac{2}{s}\left( \partial _{v}s\right) ^{2}\right) \left[ \begin{array}{ccc}
1 & 0 & 0\\
0 & 1 & 0\\
0 & 0 & 0
\end{array}\right] \\
 &  & +\frac{2}{s^{3}}\left( \left( \partial _{u}s\right) \partial _{v}-\left( \partial _{v}s\right) \partial _{u}\right) \left[ \begin{array}{ccc}
0 & -1 & 0\\
1 & 0 & 0\\
0 & 0 & 0
\end{array}\right] 
\end{array}
\end{equation}
Although the solution to (\ref{poissw}) is simple for the case of two dimensional
motion, because the \( z \) dependence is integrated out, the full three dimensional
solution is complicated by the position dependent Laplacian, and an approximation
to this solution will be made in Section \ref{sec: apgreen}.

\section{Two dimensional motion}

The goal of this analysis is to obtain a description of the behavior of a vortex
filament over a wavy surface. The dynamics of a straight filament must be determined,
and then extended via a small perturbation, to predict the filament behavior.
The behavior of the straight filament is determined by considering the motion
in the proper curvilinear coordinates. A two dimensional Hamiltonian formulation,
which arises as the limit of the full three dimensional Hamiltonian formulation,
facilitates this analysis.

The position of a straight filament, parallel to the corrugations of the surface,
is described by
\[
\overrightarrow{R^{W}_{2}}\left( t\right) =\left( u_{0}\left( t\right) ,v_{0}\left( t\right) \right) _{uv}\]
The dynamics of the filament are described via Hamiltonian dynamics, with the
fundamental Poisson bracket, arising from (\ref{pbw}), equal to
\[
\left\{ u_{0},v_{0}\right\} =\frac{1}{\Gamma s^{2}(u_{0},v_{0})}\]
and the Hamiltonian, from (\ref{hamw}), equal to
\[
\begin{array}{ccl}
H^{W}_{2} & = & \frac{\Gamma ^{2}}{2}G_{2}^{W}\left( \overrightarrow{R^{W}_{2}},\overrightarrow{R^{W}_{2}}\right) +\Gamma \Psi _{z}^{b}\\
 & = & \frac{\Gamma ^{2}}{2}\frac{1}{2\pi }\ln (2v_{0})+\Gamma Uv_{0}
\end{array}\]
in which the two dimensional Green function for the wavy surface,
\begin{equation}
\label{g2w}
\begin{array}{ccc}
G^{W}_{2}\left( \overrightarrow{R^{W}_{2}},\overrightarrow{R^{W}_{2}}'\right)  & = & -\frac{1}{2\pi }\ln \sqrt{\left( u_{0}-u_{0}'\right) ^{2}+\left( v_{0}-v_{0}'\right) ^{2}}\\
 &  & +\frac{1}{2\pi }\ln \sqrt{\left( u_{0}-u_{0}'\right) ^{2}+\left( v_{0}+v_{0}'\right) ^{2}}
\end{array}
\end{equation}
is the solution to
\begin{equation}
\label{poiss2w}
\begin{array}{ccl}
\nabla ^{2}G^{W}_{2} & = & \frac{1}{s^{2}(u_{0},v_{0})}\left( \frac{\partial ^{2}}{\partial u_{0}^{2}}+\frac{\partial ^{2}}{\partial v_{0}^{2}}\right) G^{W}_{2}\\
 & = & -\delta ^{2}(\overrightarrow{R_{2}}-\overrightarrow{R_{2}'})\\
 & = & -\frac{1}{s^{2}(u_{0},v_{0})}\delta (u_{0}-u_{0}')\delta (v_{0}-v_{0}')
\end{array}
\end{equation}
and satisfies the boundary conditions, \( \left. G_{2}^{W}\right| _{v_{0}=0}=0 \).
(Note the appearance of the scale factor, \( \frac{1}{s^{2}} \), when the delta
function in (\ref{poiss2w}) is converted from a distribution over flat to curvilinear
coordinates.) The resulting equations of motion for the straight filament propagating
over the wavy boundary are
\begin{equation}
\label{eom2w}
\begin{array}{ccccl}
\partial _{t}u_{0} & = & \left\{ u_{0},H^{W}_{2}\right\}  & = & \frac{1}{s^{2}(u_{0},v_{0})}\left( u^{b}+\frac{\Gamma }{2\pi (2v_{0})}\right) \\
\partial _{t}v_{0} & = & \left\{ v_{0},H^{W}_{2}\right\}  & = & 0
\end{array}
\end{equation}

\begin{figure}
{\centering \resizebox*{5.5in}{!}{\includegraphics{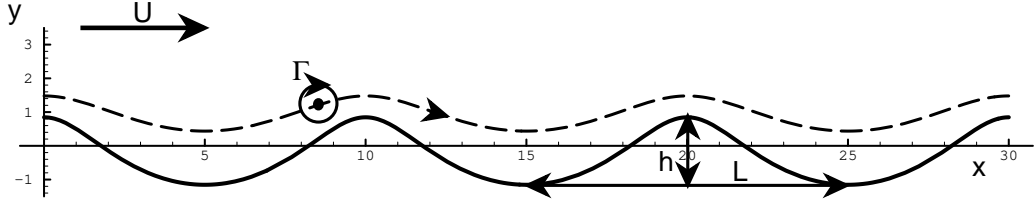}} \par}

\caption{\label{fig: fil2d} The path of a straight vortex filament at height \protect\( v_{0}=1\protect \)
propagating over a wavy boundary of wavelength \protect\( L=10\protect \) and
height to length ratio \protect\( \frac{h}{L}=.2\protect \).}
\end{figure}

The filament travels along a streamline of constant \( v_{0} \), with \( u_{0}(t) \)
steadily increasing with period \( T \) in time, \( u_{0}(nT+t)=nL+u_{0}(t) \).
The first order, nonlinear ordinary differential equation for \( u_{0}(t) \),
(\ref{eom2w}), is integrated numerically to obtain the filament position as
a function of time. The straight filament, \( \overrightarrow{R^{W}_{2}}\left( t\right) =\left( u_{0}\left( t\right) ,v_{0}\right) _{uv} \),
moving along the streamlines of constant \( v_{0} \) with varying velocity,
(Figure \ref{fig: fil2d}), is used as the basis for small three dimensional
perturbations about this solution.

It is also possible to integrate the equations of motion, (\ref{eom2w}), analytically,
rather then numerically, to obtain an equation in closed form for the filament
postion, \( u_{0} \), as a function of time. Carrying out the integration gives
\begin{equation}
\label{tu0}
\begin{array}{ccl}
t & = & \frac{1}{\left( u^{b}+\frac{\Gamma }{2\pi (2v_{0})}\right) }(-u_{0}\\
 &  & +\frac{\frac{2L}{\pi }}{\tanh \left( \frac{2\pi }{L}\left( v_{0}+v_{B}\right) \right) }\arctan \left( \tan \left( \frac{\pi }{L}u_{0}\right) \tanh \left( \frac{\pi }{L}\left( v_{0}+v_{B}\right) \right) \right) )
\end{array}
\end{equation}
which may be solved for \( u_{0}(t) \). However, solving this transcendental
equation for \( u_{0}(t) \) is no less difficult then integrating the equations
of motion numerically, and equation (\ref{tu0}) is hence used only to calculate
the period of filament motion over the surface,
\[
T=\frac{L}{\left( u^{b}+\frac{\Gamma }{2\pi (2v_{0})}\right) }\left( -1+\frac{2}{\tanh \left( \frac{2\pi }{L}\left( v_{0}+v_{B}\right) \right) }\right) \]
which is then used to set up the numerical integration of (\ref{eom2w}) over
one period.

\section{Linearized equations of three dimensional filament motion}

The Hamiltonian, (\ref{hamw}), with the Poisson brackets, (\ref{pbw}), generates
the dynamics of the vortex filament above the corrugated surface,
\begin{equation}
\label{heomw}
\begin{array}{ccccc}
\partial _{t}U(z,t) & = & \left\{ U,H^{W}\right\}  & = & \frac{1}{\Gamma s^{2}(U,V)}\frac{\delta H^{W}}{\delta V}\\
\partial _{t}V(z,t) & = & \left\{ V,H^{W}\right\}  & = & -\frac{1}{\Gamma s^{2}(U,V)}\frac{\delta H^{W}}{\delta U}
\end{array}
\end{equation}
These equations of motion are linearized about the straight filament solution,
as was done in Section \ref{sec: linearization}, with the breakup of \( U \)
and \( V \) into two dimensional and small three dimensional terms,
\begin{equation}
\label{pertw}
\begin{array}{ccl}
U\left( z,t\right)  & = & u_{0}\left( t\right) +\varepsilon \widetilde{u}\left( z,t\right) \\
V\left( z,t\right)  & = & v_{0}+\varepsilon \widetilde{v}\left( z,t\right) 
\end{array}
\end{equation}
The Hamiltonian is expanded order by order in \( \varepsilon  \) to obtain
\begin{equation}
H^{W}=H^{W\left( 0\right) }+\varepsilon H^{W\left( 1\right) }+\varepsilon ^{2}H^{W\left( 2\right) }+O(\varepsilon ^{3})
\end{equation}
The equations of motion, (\ref{heomw}), are then similarly expanded order by
order in \( \varepsilon  \), with the first order equations producing the linearized
equations of motion for the filament,
\begin{equation}
\label{leomw}
\begin{array}{ccl}
\partial _{t}\widetilde{u} & = & -\frac{\partial _{t}u_{o}}{s^{2}}\left( \frac{\partial s^{2}}{\partial u}\widetilde{u}+\frac{\partial s^{2}}{\partial v}\widetilde{v}\right) +\frac{1}{\Gamma s^{2}}\frac{\delta H^{W\left( 2\right) }}{\delta \widetilde{v}}\\
\partial _{t}\widetilde{v} & = & -\frac{1}{\Gamma s^{2}}\frac{\delta H^{W\left( 2\right) }}{\delta \widetilde{u}}
\end{array}
\end{equation}
in which \( s^{2} \) and its derivatives are evaluated at the filament center,
\( \left( u_{0},v_{0}\right)  \). The quadratic part of the Hamiltonian is
equal to
\begin{equation}
\label{qhamw}
H^{W\left( 2\right) }=\frac{\Gamma ^{2}}{2}\int \int dzdz'\left( \frac{\partial \widetilde{R_{\alpha }}}{\partial z}G_{\alpha \beta }^{W\left( 0\right) }\frac{\partial \widetilde{R_{\beta }}}{\partial z'}+G_{zz}^{W\left( 2\right) }\right) 
\end{equation}
in which \( \widetilde{\overrightarrow{R}}=\left( \widetilde{u},\widetilde{v}\right)  \),
Greek indices \( \left\{ \alpha ,\beta \right\}  \) range only over \( \left\{ u,v\right\}  \),
and \( G^{W\left( 0\right) }_{\alpha \beta } \) and \( G^{W\left( 2\right) }_{zz} \)
are the modified Green function components of zeroth and quadratic order in
\( \varepsilon  \),
\[
\begin{array}{ccl}
G^{W\left( 0\right) }_{\alpha \beta } & = & \left. G^{W}_{\alpha \beta }(\overrightarrow{R}(z),\overrightarrow{R}(z'))\right| _{\varepsilon =0}\\
G^{W\left( 0\right) }_{zz} & = & \left. \frac{1}{2}\frac{\partial ^{2}}{\partial \varepsilon ^{2}}G^{W}_{zz}(\overrightarrow{R}(z),\overrightarrow{R}(z'))\right| _{\varepsilon =0}
\end{array}\]

\section{The approximate Green function\label{sec: apgreen} }

The calculation of the linear equations of motion for the vortex filament over
the wavy boundary, (\ref{leomw}), requires the use of the Green function solution
to the singular Poisson equation, (\ref{poissw}), satisfying the boundary conditions,
(\ref{bcgws}). This Green function may not be easily found in closed form because
of the non-trivial nature of the position dependent Laplace operator in the
wavy coordinates, (\ref{lapw}). However, it is possible to approximate this
Green function, and to improve upon this approximation by employing a Neumann
series.

The approximate Green function is required to have the following properties:
it must reduce to the flat boundary Green function as \( \frac{h}{L}\rightarrow 0 \),
it must integrate to the correct two dimensional Green function, (\ref{g2w}),
and it must satisfy the boundary conditions at the surface, (\ref{bcgws}).
The flat boundary Green function, (\ref{sgflat}), with curvilinear coordinates
as arguments,
\begin{equation}
\label{gfw}
\begin{array}{ccl}
G^{W} & \simeq  & G^{F}\left( (u,v,z),(u',v',z')\right) \\
 & = & \frac{1}{4\pi \sqrt{\left( u-u'\right) ^{2}+\left( v-v'\right) ^{2}+\left( z-z'\right) ^{2}+e^{-\frac{3}{2}}a^{2}}}\left[ \begin{array}{ccc}
1 & 0 & 0\\
0 & 1 & 0\\
0 & 0 & 1
\end{array}\right] \\
 &  & -\frac{1}{4\pi \sqrt{\left( u-u'\right) ^{2}+\left( v+v'\right) ^{2}+\left( z-z'\right) ^{2}}}\left[ \begin{array}{ccc}
1 & 0 & 0\\
0 & -1 & 0\\
0 & 0 & 1
\end{array}\right] 
\end{array}
\end{equation}
satisfies these conditions. It is a good approximation to the exact Green function
for the wavy boundary when applied in the case of small amplitude filament perturbations
and small \( \frac{h}{L} \).

Although the approximate Green function, (\ref{gfw}), is not an exact solution
to the Poisson equation, it is an approximate solution, in that it is a solution
to
\begin{equation}
\label{poissf}
\frac{1}{s^{2}}\nabla ^{2F}G^{F}_{ij}\left( (u,v,z),(u',v',z')\right) =-\frac{1}{s^{2}}\delta (u-u')\delta (v-v')\delta (z-z')\delta _{ij}
\end{equation}
with the ``flat'' Laplace operator, \( \nabla ^{2F}=\partial ^{2}_{u}+\partial ^{2}_{v}+\partial ^{2}_{z} \).
Equation (\ref{poissf}) approximates the Poisson equation, (\ref{poissw}),
which may be written as
\begin{equation}
\label{poissww}
\begin{array}{ccl}
\nabla _{ik}^{2}G^{W}_{kj}\left( (u,v,z),(u',v',z')\right)  & = & \frac{1}{s^{2}}\left( \nabla ^{2F}\delta _{ik}+\nabla _{ik}^{2W}\right) G^{W}_{kj}\\
 & = & -\frac{1}{s^{2}}\delta (u-u')\delta (v-v')\delta (z-z')\delta _{ij}\\
 & = & -\delta ^{3}(\overrightarrow{x}-\overrightarrow{x'})\delta _{ij}
\end{array}
\end{equation}
by breaking the Laplace operator in curvilinear coordinates, (\ref{lapw}),
into a ``flat'' part, \( \nabla ^{2F} \), and ``wavy'' part,

\begin{equation}
\label{lapww}
\begin{array}{ccl}
\nabla ^{2W} & = & s^{2}\nabla ^{2}-\nabla ^{2F}\\
 & = & \left( s^{2}-1\right) \partial ^{2}_{z}\left[ \begin{array}{ccc}
1 & 0 & 0\\
0 & 1 & 0\\
0 & 0 & 1
\end{array}\right] \\
 &  & +\frac{1}{s}\left( \left( \partial ^{2}_{u}s\right) -\frac{2}{s}\left( \partial _{u}s\right) ^{2}+\left( \partial ^{2}_{v}s\right) -\frac{2}{s}\left( \partial _{v}s\right) ^{2}\right) \left[ \begin{array}{ccc}
1 & 0 & 0\\
0 & 1 & 0\\
0 & 0 & 0
\end{array}\right] \\
 &  & +\frac{2}{s}\left( \left( \partial _{u}s\right) \partial _{v}-\left( \partial _{v}s\right) \partial _{u}\right) \left[ \begin{array}{ccc}
0 & -1 & 0\\
1 & 0 & 0\\
0 & 0 & 0
\end{array}\right] 
\end{array}
\end{equation}
which goes to \( 0 \) as \( \frac{h}{L}\rightarrow 0 \).

\subsection{Neumann series}

If necessary, the approximate Green function, (\ref{gfw}), may be improved
by adding terms from a Neumann series expansion. By rewriting (\ref{poissww})
as
\begin{equation}
\label{fpoissw}
\nabla ^{2F}G_{ij}^{W}=-\delta (u-u')\delta (v-v')\delta (z-z')\delta _{ij}-\nabla _{ik}^{2W}G_{kj}^{W}=-\rho _{ij}(u,v,z)
\end{equation}
considering \( \rho  \) to be the source for a ``flat'' Poisson equation,
and using the ``flat'' Green function, \( G^{F} \), to ``solve'' for the
\( G^{W} \) on the left hand side of (\ref{fpoissw}), a Fredholm equation
of the second kind for \( G^{W} \) is obtained,
\[
\begin{array}{ccl}
G_{ij}^{W} & = & \int \int \int du''dv''dz''G_{ik}^{F}\left( (u,v,z),(u'',v'',z'')\right) \rho _{kj}(u'',v'',z'')\\
 & = & G_{ij}^{F}+\int \int \int du''dv''dz''G_{ik}^{F}\nabla _{kl}^{2W}G_{lj}^{W}
\end{array}\]
This integral equation may be solved for \( G^{W} \) by recursive iteration,
producing a Neumann series,
\begin{equation}
\label{neumann}
G_{ij}^{W}=G_{ij}^{F}+\int \int \int du''dv''dz''G_{ik}^{F}\nabla _{kl}^{2W}G_{lj}^{F}+...
\end{equation}
which will converge for small \( \nabla ^{2W} \), and hence for small \( \frac{h}{L} \).

In practice, the integrals in (\ref{neumann}) cannot be written in closed form,
and must be computed numerically to determine the corrections to the Green function.
This procedure is not necessary to the calculation of the linear instability
of the filament over a surface of small \( \frac{h}{L} \), which uses only
the simple approximate Green function, \( G^{W}\simeq G^{F}\left( (u,v,z),(u',v',z')\right)  \).

\section{Fourier transform and equations of motion}

The diagonalization of (\ref{qhamw}) is carried out via a Fourier transform
in the \( z \) coordinate, as in Section \ref{sec: diag}, with the small dynamical
variables of the perturbation becoming the Fourier transformed pair, \( \widetilde{u}(k,t) \)
and \( \widetilde{v}(k,t) \). The use of the simple approximate Green function
produces an approximate quadratic Hamiltonian identical to (\ref{fham}),
\[
H^{W\left( 2\right) }\simeq \frac{\Gamma ^{2}}{2}\int dk\, k^{2}\left\{ A^{W}(k)\, \widetilde{u}^{*}(k,t)\, \widetilde{u}\left( k,t\right) +B^{W}(k)\, \widetilde{v}^{*}(k,t)\, \widetilde{v}(k,t)\right\} \]
with \( A \) and \( B \) given, as before, by
\[
\begin{array}{ccl}
A^{W} & = & S(ka)+\left\{ \frac{1}{\left( 2kv_{0}\right) ^{2}}-K_{0}(2kv_{0})-\frac{1}{2kv_{0}}K_{1}(2kv_{0})\right\} \\
B^{W} & = & S(ka)+\left\{ -\frac{1}{\left( 2kv_{0}\right) ^{2}}-\frac{1}{2kv_{0}}K_{1}(2kv_{0})\right\} 
\end{array}\]
with the self interaction term given by (\ref{Ska}).

The linear equations of motion, (\ref{leomw}), for each mode then become, in
matrix form,
\begin{equation}
\label{lfeomw}
\partial _{t}\left[ \begin{array}{c}
\widetilde{u}\\
\widetilde{v}
\end{array}\right] =\frac{1}{s^{2}}\left[ \begin{array}{cc}
-\left( \partial _{t}u_{o}\right) \frac{\partial s^{2}}{\partial u} & -\left( \partial _{t}u_{o}\right) \frac{\partial s^{2}}{\partial v}+\frac{\Gamma k^{2}}{2\pi }B^{W}\\
-\frac{\Gamma k^{2}}{2\pi }A^{W} & 0
\end{array}\right] \left[ \begin{array}{c}
\widetilde{u}\\
\widetilde{v}
\end{array}\right] 
\end{equation}
with \( s^{2} \) and its derivatives evaluated at \( (u_{0}(t),v_{0}) \).
The effect of the wavy boundary is apparent in the difference of (\ref{lfeomw})
from the equations of motion in the presence of a flat boundary, (\ref{lfeom}).
The most important difference is the appearance of time dependent terms in the
matrix coefficients, due to the periodic motion of the straight vortex filament
over the wavy surface, \( u_{0}(t) \). This affects the filament primarily
through an overall time periodic amplitude stretching, \( \frac{1}{s^{2}(u_{0}(t),v_{0})} \),
due to the expansion and contraction of the flow streamlines over the surface.
The perturbative mode amplitudes, \( \widetilde{u} \) and \( \widetilde{v} \),
are also affected independently by the changes in the flow velocity field surrounding
the propagating filament -- a result of the velocity field, and hence the \( uv \)
frame, rocking rapidly back and forth as the center of the filament travels
over the undulations.

Although the method of images is not accurate for a wavy surface, it is a good
approximation when the filament height is small compared to the period, \( \frac{h}{L}\ll 1 \),
and provides an alternative view of the effects of the wavy surface on the filament
instability. As the filament center travels over the surface, following a streamline,
(Figure \ref{fig: fil2d}), it travels periodically closer to and farther from
the boundary, with its image filament drawing correspondingly nearer and farther.
This rapid, time periodic, change in the most fundamental parameter, the vortex
filament height from the surface, has a significant effect on the evolution
of the vortex filament perturbation amplitudes.

\section{Floquet analysis }

The linear equations of motion, (\ref{lfeomw}), governing the evolution of
the vortex filament perturbation for each wavenumber, \( k \), contain coefficients
that are periodic in time, with period \( T \). Floquet's theorem implies that
the general solution to these equations may be written as
\begin{equation}
\label{flosol}
\left[ \begin{array}{c}
\widetilde{u}(t)\\
\widetilde{v}(k)
\end{array}\right] =ae^{\sigma _{1}t}\left[ \begin{array}{c}
f_{1}\left( t\right) \\
f_{2}\left( t\right) 
\end{array}\right] +be^{\sigma _{2}t}\left[ \begin{array}{c}
g_{1}\left( t\right) \\
g_{2}\left( t\right) 
\end{array}\right] 
\end{equation}
with complex Floquet exponents, \( \left\{ \sigma _{1},\sigma _{2}\right\}  \),
time periodic functions, \( \left\{ f_{1},f_{2},g_{1},g_{2}\right\}  \), and
coefficients , \( \left\{ a,b\right\}  \), determined by the initial conditions.

An inspection of the time reversal symmetries of the coefficients in the equations
of motion provides further information about the solution, allowing the second
Floquet solution set, \( \{g_{1},g_{2}\} \) and \( \sigma _{2} \), to be related
to the first. The symmetries of the coefficients in (\ref{lfeomw}) are
\[
\begin{array}{rcl}
s^{2}\left( -t\right)  & = & s^{2}\left( t\right) \\
\left. \partial _{t}u_{0}\right| _{-t} & = & \left. \partial _{t}u_{0}\right| _{t}\\
\left. \frac{\partial s^{2}}{\partial u}\right| _{-t} & = & -\left. \frac{\partial s^{2}}{\partial u}\right| _{t}\\
\left. \frac{\partial s^{2}}{\partial v}\right| _{-t} & = & \left. \frac{\partial s^{2}}{\partial v}\right| _{t}
\end{array}\]
Hence, the existence of a solution with the time dependence
\[
\left[ \begin{array}{c}
\widetilde{u}\\
\widetilde{v}
\end{array}\right] =e^{\sigma _{1}t}\left[ \begin{array}{c}
f_{1}\left( t\right) \\
f_{2}\left( t\right) 
\end{array}\right] \]
implies -- through time reversal of the equations of motion -- the existence
of the solution
\[
\left[ \begin{array}{c}
\tilde{u}\\
\tilde{v}
\end{array}\right] =e^{-\sigma _{1}t}\left[ \begin{array}{c}
f_{1}\left( -t\right) \\
-f_{2}\left( -t\right) 
\end{array}\right] \]
Since the general solution, (\ref{flosol}), consists of only two independent
terms, the second term must accommodate this time mirrored solution, and hence
\begin{equation}
\label{flosym}
\begin{array}{rcl}
g_{1}\left( t\right)  & = & f_{1}\left( -t\right) \\
g_{2}\left( t\right)  & = & -f_{2}\left( -t\right) \\
\sigma _{2} & = & -\sigma _{1}
\end{array}
\end{equation}
It is not surprising that the Floquet exponents are inverses, since the motion
is due to Hamiltonian flow and to coordinate frame rotation.

The stability of the filament is now completely described by the Floquet solution.
Since the exponents are inverses, the perturbation has either neutral stability,
with \( \Re (\sigma _{1})=0 \), or a pair of unstable and stable modes, with
exponents \( \sigma _{\pm }\equiv \pm \left| \sigma _{1}\right|  \)and \( \Im (\sigma _{1})=0 \).
A complete description of the linear behavior of the perturbation to the vortex
filament requires only the calculation of \( f_{1}(t) \), \( f_{2}(t) \),
and \( \sigma _{1} \).

\section{Numerical implementation }

Although analytic methods exist to calculate the Floquet exponent, \( \sigma _{1} \),
they are not practical for the current case because of the complicated time
dependent coefficients in the equations of motion, (\ref{lfeomw}). Rather,
a numerical solution over one period, \( T \), of the motion is used to determine
the Floquet functions, \( f_{1} \) and \( f_{2} \), and Floquet exponent,
\( \sigma _{1} \), as a function of wave number, \( k \), filament height,
\( v_{0} \), background flow, \( U \), and surface geometry, \( h \) and
\( L \).

Of primary interest are the size of the growth parameter, \( \sigma _{+} \),
if non-zero, and the angle, \( \theta =\arctan (\frac{f_{2}}{f_{1}}) \), at
which the perturbation grows over one period. To determine these quantities,
the equations of motion, (\ref{lfeomw}), are numerically integrated over one
period of motion from the initial conditions,
\[
\begin{array}{ccc}
\widetilde{u}(t=0) & = & 1\\
\widetilde{v}(t=0) & = & 0
\end{array}\]
to obtain the resulting perturbation amplitudes at the end of one period of
motion over the surface, \( \widetilde{u}(T) \) and \( \widetilde{v}(T) \).
These quantities are compared with the general Floquet solution, (\ref{flosol}),
to obtain the equations,
\begin{equation}
\label{flonum}
\begin{array}{rcl}
1 & = & af_{1}+bg_{1}\\
0 & = & af_{2}+bg_{2}\\
\widetilde{u}(T) & = & a\lambda _{1}f_{1}+b\lambda _{2}g_{1}\\
\widetilde{v}(T) & = & a\lambda _{1}f_{2}+b\lambda _{2}g_{2}
\end{array}
\end{equation}
in which \( \lambda _{1}=e^{\sigma _{1}T} \) and \( \lambda _{2}=e^{\sigma _{2}T} \)
are the Floquet multipliers, and the Floquet functions are evaluated at \( t=0 \)
(or \( t=T \)). The time symmetry relations, (\ref{flosym}), are used to eliminate
\( g_{1} \), \( g_{2} \), and \( \lambda _{2} \) in (\ref{flonum}), and
the resulting equations are solved for \( \lambda _{1} \) and \( \frac{f_{2}}{f_{1}} \),
\[
\begin{array}{rcl}
\lambda _{1,2} & = & \widetilde{u}(T)\pm \sqrt{\widetilde{u}(T)^{2}-1}\\
\frac{f_{2}}{f_{1}} & = & \frac{\widetilde{v}(T)}{\sqrt{\widetilde{u}(T)^{2}-1}}
\end{array}\]
The filament is linearly unstable to perturbations if and only if \( \left| \widetilde{u}(T)\right| >1 \),
with the growth parameter,
\[
\sigma _{+}=\frac{1}{T}\ln \left( \left| \widetilde{u}(T)\right| +\sqrt{\widetilde{u}(T)^{2}-1}\right) \]
If \( \widetilde{u}(T)>1 \), the angle of this unstable mode, after each period,
\( T \), is
\[
\theta _{+}=\arctan \left( \frac{\widetilde{v}(T)}{\sqrt{\widetilde{u}(T)^{2}-1}}\right) \]
and if \( \widetilde{u}(T)<-1 \), 
\[
\theta _{+}=\arctan \left( \frac{-\widetilde{v}(T)}{\sqrt{\widetilde{u}(T)^{2}-1}}\right) \]
If \( \left| \widetilde{u}(T)\right| <1 \), the Floquet exponents are pure
imaginary and the filament has neutral stability, with perturbations at this
wavenumber oscillating, but not growing, about the filament center as it propagates
over the surface.

\section{Stability results}

\begin{figure}
{\centering \resizebox*{4.5in}{!}{\includegraphics{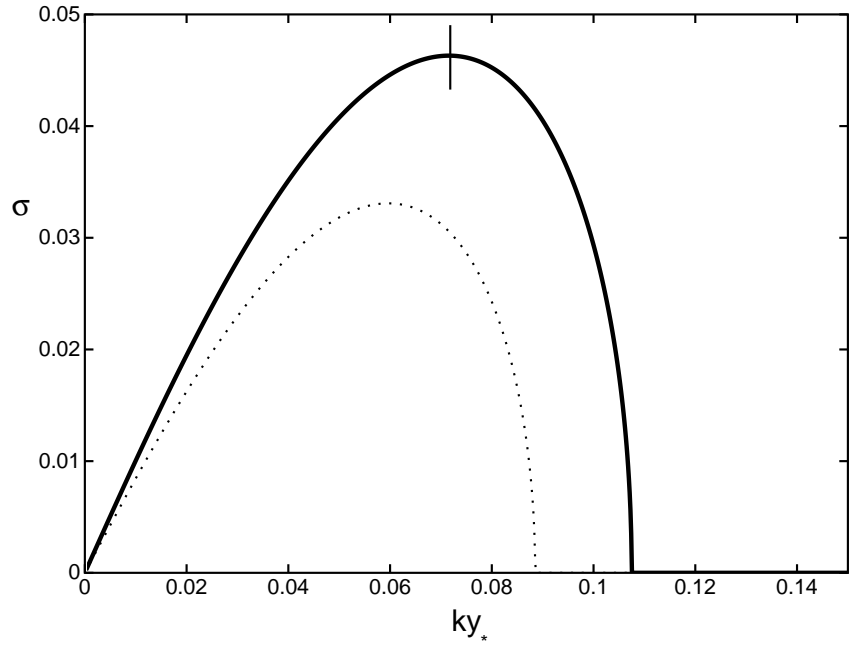}} \par}

\caption{\label{fig: stabw}The growth parameter, \protect\( \sigma _{+}\protect \),
in units of \protect\( \frac{\left| \Gamma \right| }{v_{0}^{2}}\protect \),
as a function of spanwise wavenumber, \protect\( ky_{*}\protect \), for a filament
over a wavy boundary of length \protect\( L=0.61mm\protect \), and height to
length ratio, \protect\( \frac{h}{L}=0.1\protect \). The maximum at \protect\( ky_{*}\simeq 0.072\protect \)
is marked. The free stream velocity is \protect\( U=3\frac{m}{s}\protect \).
The growth parameter for the flat plate is shown for comparison, as a dotted
curve.}
\end{figure}

\begin{figure}
{\centering \resizebox*{4.5in}{!}{\includegraphics{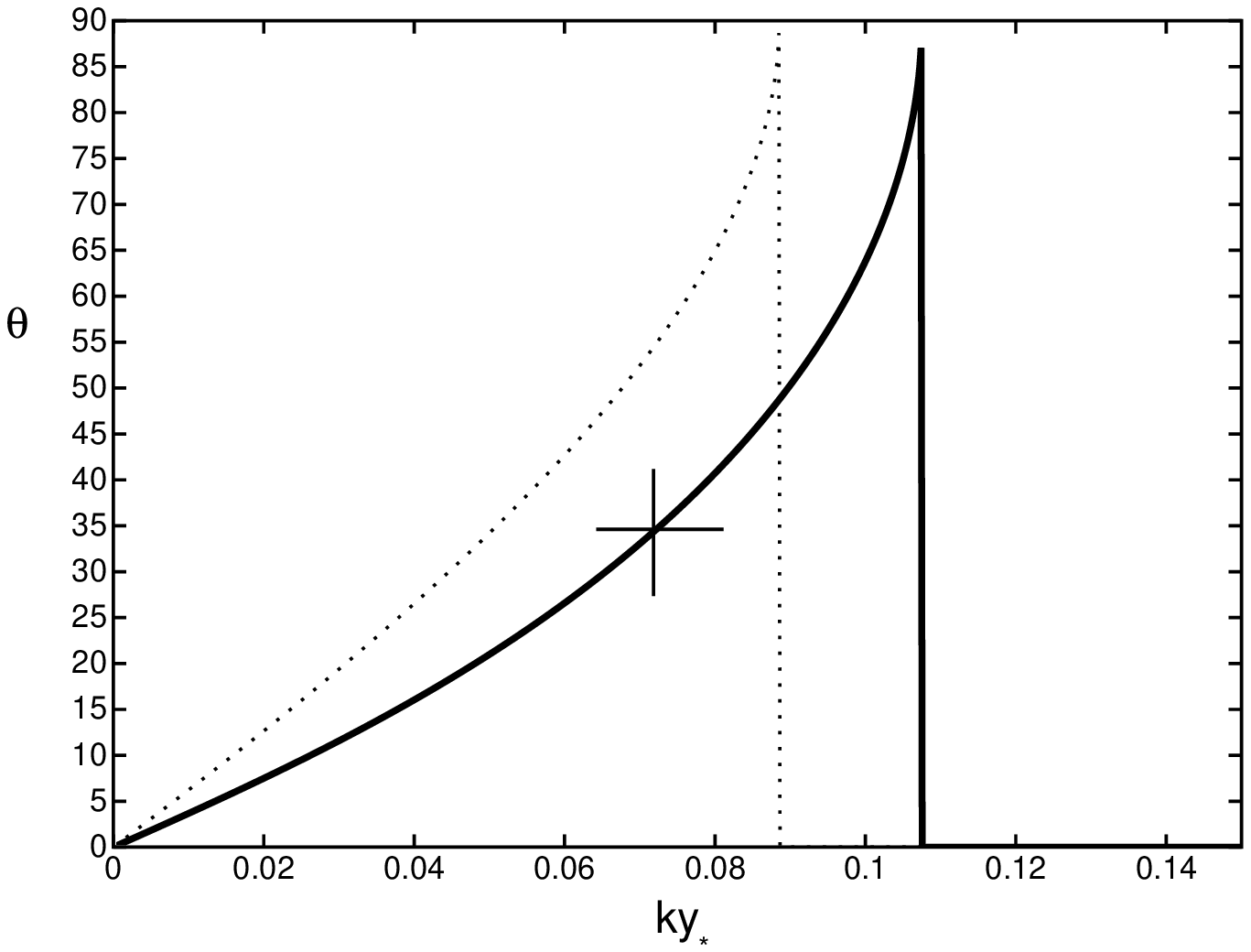}} \par}

\caption{\label{fig: angw}The instability angle, \protect\( \theta _{+}\protect \),
as a function of spanwise wavenumber, \protect\( ky_{*}\protect \), for a filament
over a wavy boundary of length \protect\( L=0.61mm\protect \), and height to
length ratio, \protect\( \frac{h}{L}=0.1\protect \). The maximally unstable
wavenumber, \protect\( ky_{*}\simeq 0.072\protect \), is marked, indicating
the angle of the maximally unstable mode, \protect\( \theta _{+}\simeq 34^{o}\protect \).
The instability angle for the flat plate is shown as a dotted curve.}
\end{figure}

The wavy boundary significantly affects the evolution of the vortex filament.
As a representative example, a plot of the growth parameter for a filament propagating
over a boundary with height to length ratio \( \frac{h}{L}=.1 \) is shown in
Figure \ref{fig: stabw}. Previously stable wavenumbers become unstable, and
the wavenumber and growth parameter of the maximally unstable mode increase.
This indicates that the hairpin vortices evolving over a wavy boundary will
be more tightly packed, having characteristic spanwise wavelengths less then
\( 100wu \). The wavy boundary also has a significant effect on the growth
angle of the hairpin vortices. The angle of the unstable modes, corresponding to the plot in Figure \ref{fig: stabw},
is shown in Figure \ref{fig: angw}. The maximally unstable mode over this wavy
boundary now grows at an angle of approximately \( 34^{o} \), a significant
change to the \( 43^{o} \) growth over a flat plate.

\begin{figure}
{\centering \resizebox*{4.5in}{!}{\includegraphics{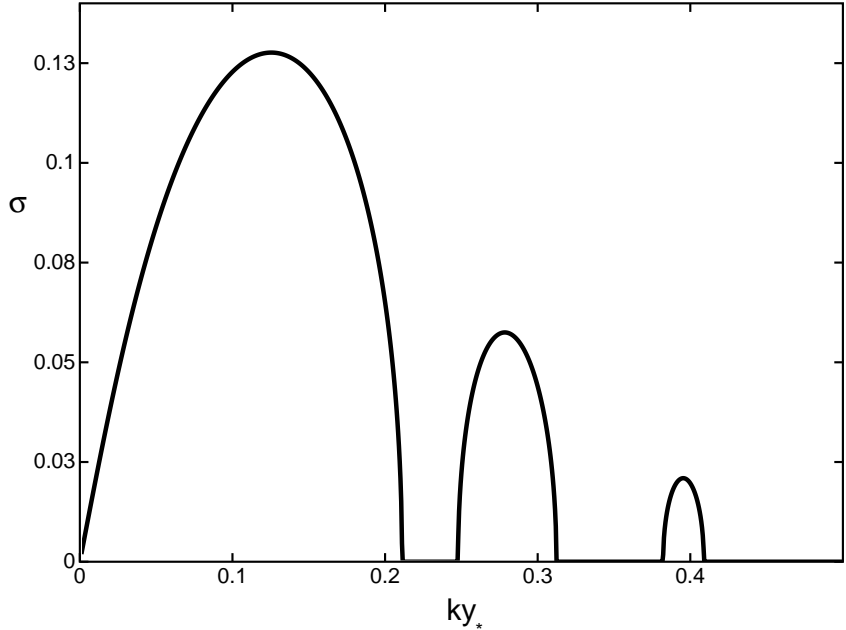}} \par}

\caption{\label{fig: band}The growth parameter, \protect\( \sigma _{+}\protect \),
in units of \protect\( \frac{\left| \Gamma \right| }{v_{0}^{2}}\protect \),
as a function of spanwise wavenumber, \protect\( ky_{*}\protect \), for a filament
over a wavy boundary of length \protect\( L=0.61mm\protect \), and height to
length ratio, \protect\( \frac{h}{L}=0.4\protect \). The free stream velocity
is \protect\( U=3\frac{m}{s}\protect \). }
\end{figure}

\begin{figure}
{\centering \resizebox*{4.5in}{!}{\includegraphics{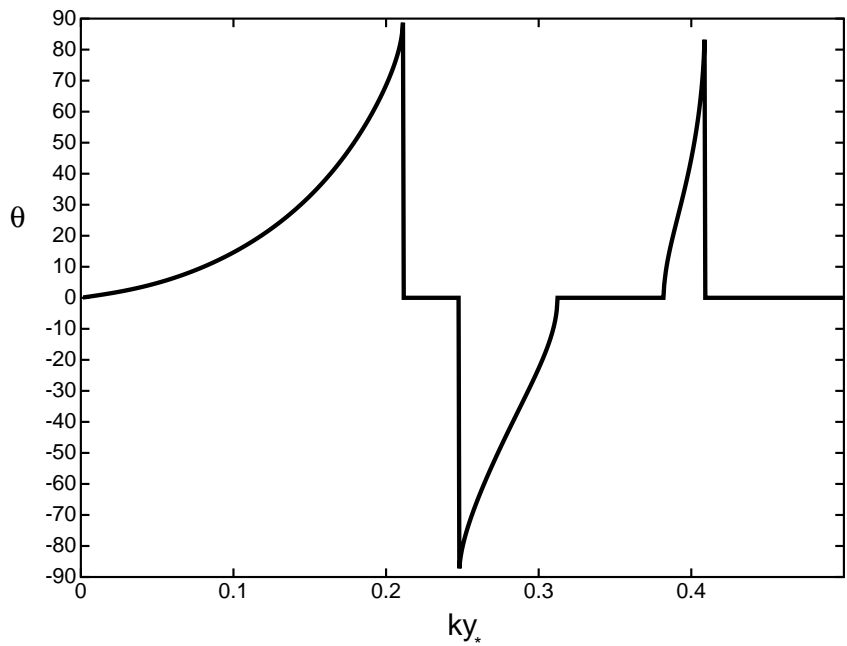}} \par}

\caption{\label{fig: banda}The instability angle, \protect\( \theta _{+}\protect \),
as a function of spanwise wavenumber, \protect\( ky_{*}\protect \), for a filament
over a wavy boundary of length \protect\( L=0.61mm\protect \), and height to
length ratio, \protect\( \frac{h}{L}=0.4\protect \). }
\end{figure}

For very large values of \( \frac{h}{L} \), the strong forcing is sufficient
to induce new instability bands at previously stable wavenumbers, (Figure \ref{fig: band}).
Although this new instability region has a smaller growth parameter than the
primary instability, and will hence not significantly effect the evolution of
a randomly perturbed vortex filament, it is worth noting that the instability
angle of the first of these new instability bands has the opposite sign of the
previous instability, (Figure \ref{fig: banda}). A filament perturbed at this
wavenumber grows in the previously stable direction, between \( -90^{o} \)
and \( 0^{o} \), illustrating the significant dynamical effect of the surface
ridges.

\begin{figure}
{\centering \includegraphics{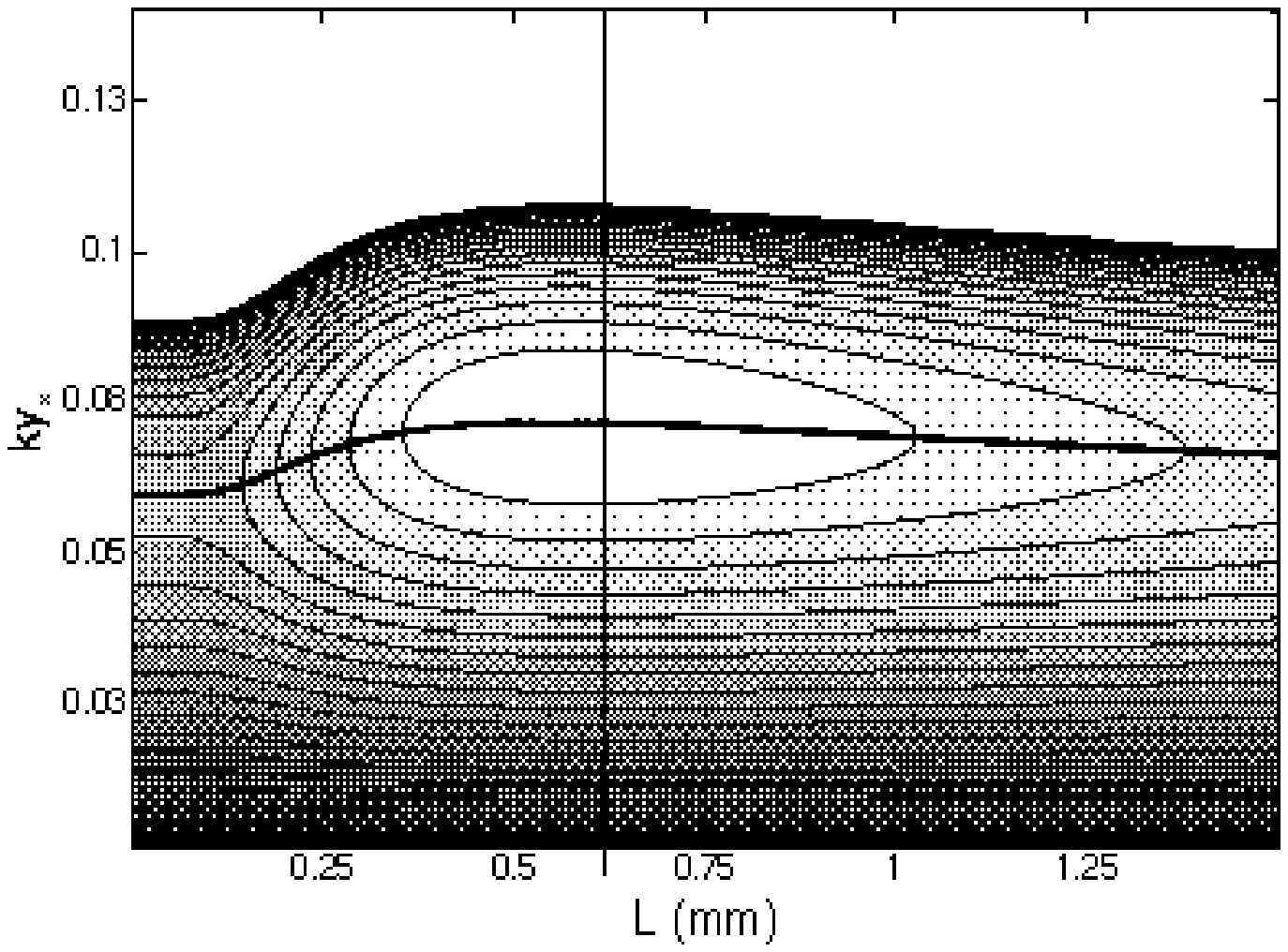} \par}

\caption{\label{fig: egg}A contour plot of the growth parameter, \protect\( \frac{\sigma _{+}v_{0}^{2}}{\left| \Gamma \right| }\protect \),
as a function of spanwise wavenumber, \protect\( ky_{*}\protect \), and corrugation
wavelength, \protect\( L\protect \), for a filament in background flow, \protect\( U=3\frac{m}{s}\protect \),
over a wavy boundary of height to length ratio, \protect\( \frac{h}{L}=0.1\protect \).
The corrugation wavelength for a dolphin, \protect\( L\simeq .61mm\protect \),
is marked by a vertical line. The maximally unstable wavenumber, \protect\( k_{+}y_{*}\protect \),
is plotted over the graph as a function of \protect\( L\protect \).}
\end{figure}

\begin{figure}
{\centering \resizebox*{4.5in}{!}{\includegraphics{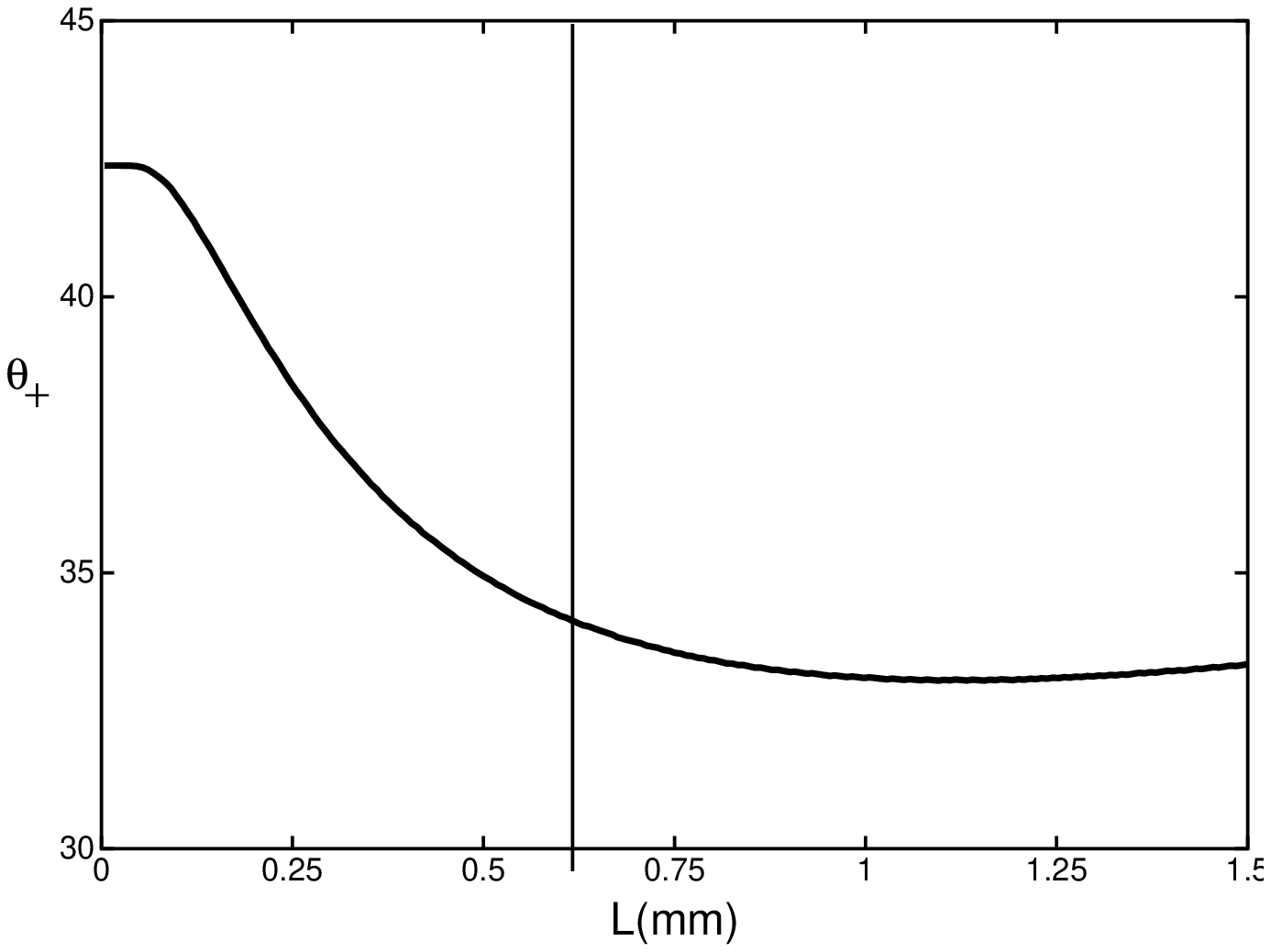}} \par}

\caption{\label{fig: thetap}A plot of the angle of the maximally unstable mode, \protect\( \theta _{+}\protect \),
as a function of corrugation wavelength, \protect\( L\protect \), for a filament
in background flow, \protect\( U=3\frac{m}{s}\protect \), over a wavy boundary
of height to length ratio, \protect\( \frac{h}{L}=0.1\protect \). The corrugation
wavelength for a dolphin, \protect\( L\simeq .61mm\protect \), is marked by
a vertical line.}
\end{figure}

A contour plot of the growth parameter for a range of surface corrugation wavelengths,
\( L \), is shown in Figure \ref{fig: egg}, and a plot of the maximal growth
angle is shown in Figure \ref{fig: thetap}. If the corrugations are small compared
to the filament height, \( L\ll v_{0} \), the surface is effectively smooth
and does not affect the filament dynamics. Alternatively, if the corrugation
length is very large, \( L\gg v_{0} \), the parametric oscillations are too
slow to affect the filament dynamics significantly. However, when the corrugation
length properly matches the filament height, \( L\simeq 20v_{0} \), the dynamics
of the filament are maximally affected, with an increase in maximally unstable
wavenumber and a decrease in instability angle for the vortex filament. The
surface corrugation wavelength of the dolphin, at \( L\simeq .61mm \), is well
tuned to the corresponding height of vortex filament formation for a large range
of swimming velocities. 

\begin{figure}
{\centering \includegraphics{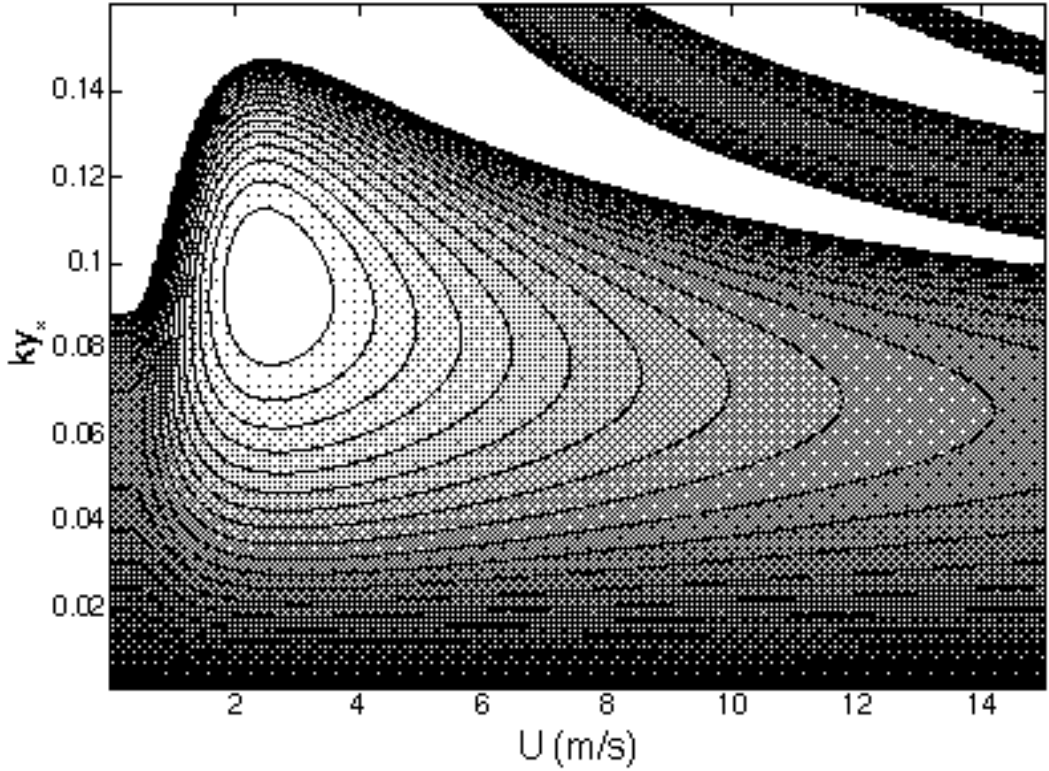} \par}

\caption{\label{fig: stab.2} The growth parameter, \protect\( \frac{\sigma _{+}v_{0}^{2}}{\left| \Gamma \right| }\protect \),
as a function of spanwise wavenumber, \protect\( ky_{*}\protect \), and background
velocity, \protect\( U\protect \), for a filament at height \protect\( v_{0}=6wu\protect \)
propagating over a wavy boundary of height to length ratio \protect\( \frac{h}{L}=0.2\protect \)
and length \protect\( L=.61mm\protect \). }
\end{figure}

\begin{figure}
{\centering \resizebox*{4.5in}{!}{\includegraphics{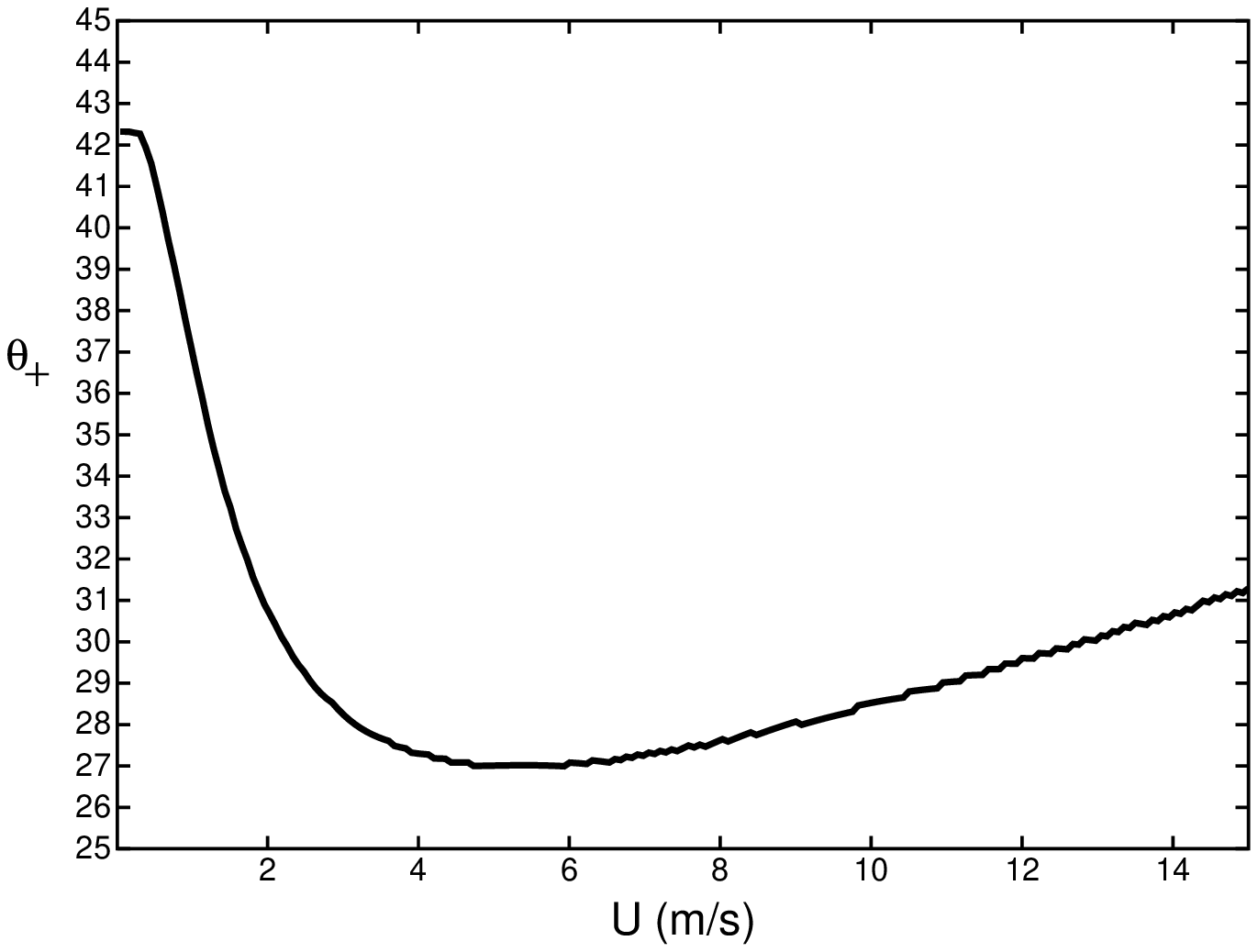}} \par}

\caption{\label{fig: staba.2}The angle of the maximally unstable mode, \protect\( \theta _{+}\protect \),
as a function of background velocity, \protect\( U\protect \), for a filament
at height \protect\( v_{0}=6wu\protect \) propagating over a wavy boundary
of height to length ratio \protect\( \frac{h}{L}=0.2\protect \) and length
\protect\( L=.61mm\protect \).}
\end{figure}

\begin{figure}
{\centering \includegraphics{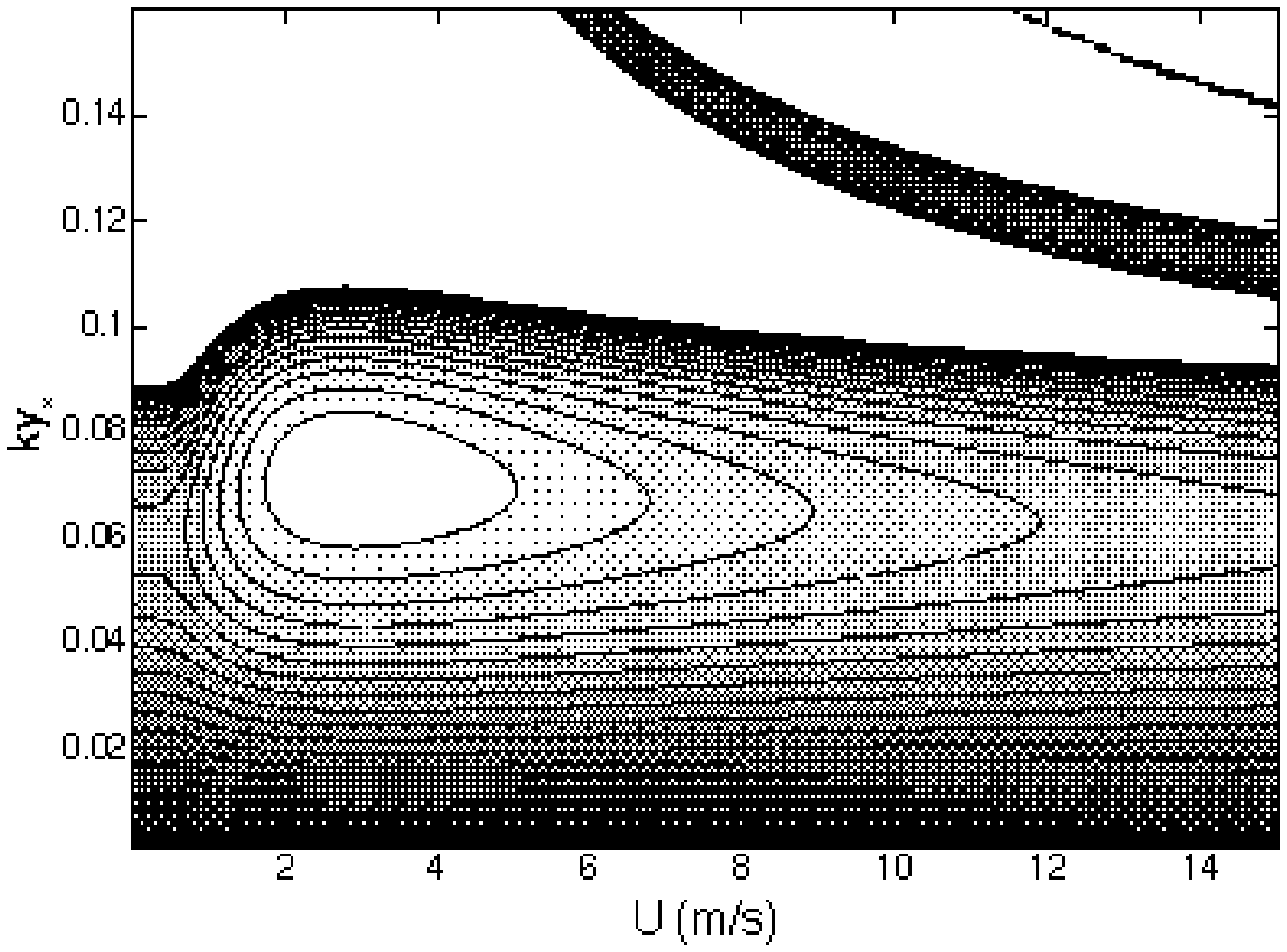} \par}

\caption{\label{fig: stab.1}The growth parameter, \protect\( \frac{\sigma _{+}v_{0}^{2}}{\left| \Gamma \right| }\protect \),
as a function of spanwise wavenumber, \protect\( ky_{*}\protect \), and background
velocity, \protect\( U\protect \), for a filament at height \protect\( v_{0}=6wu\protect \)
propagating over a wavy boundary of height to length ratio \protect\( \frac{h}{L}=0.1\protect \)
and length \protect\( L=.61mm\protect \). }
\end{figure}

\begin{figure}
{\centering \resizebox*{4.5in}{!}{\includegraphics{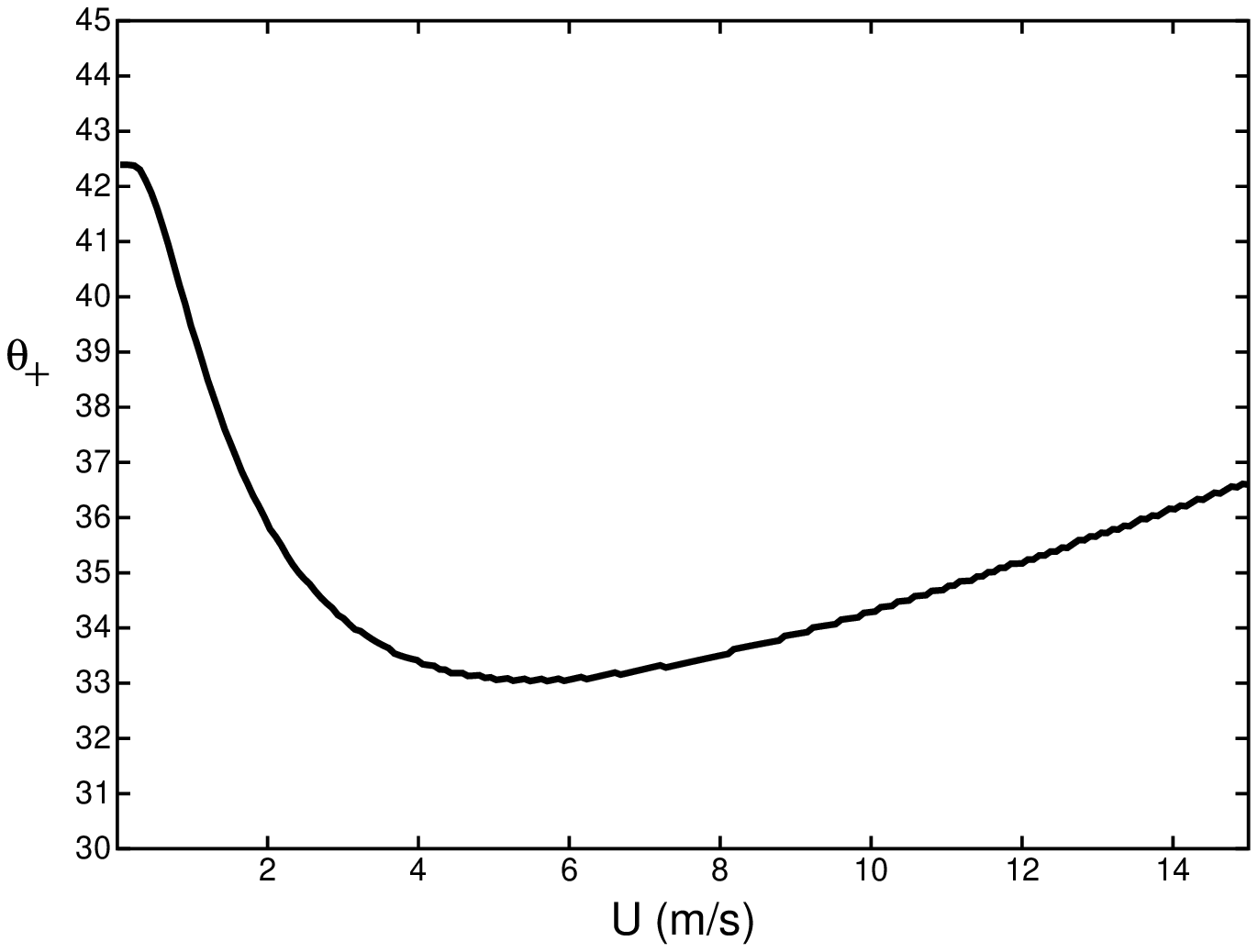}} \par}

\caption{\label{fig: staba.1}The angle of the maximally unstable mode, \protect\( \theta _{+}\protect \),
as a function of background velocity, \protect\( U\protect \), for a filament
at height \protect\( v_{0}=6wu\protect \) propagating over a wavy boundary
of height to length ratio \protect\( \frac{h}{L}=0.1\protect \) and length
\protect\( L=.61mm\protect \).}
\end{figure}

\begin{figure}
{\centering \includegraphics{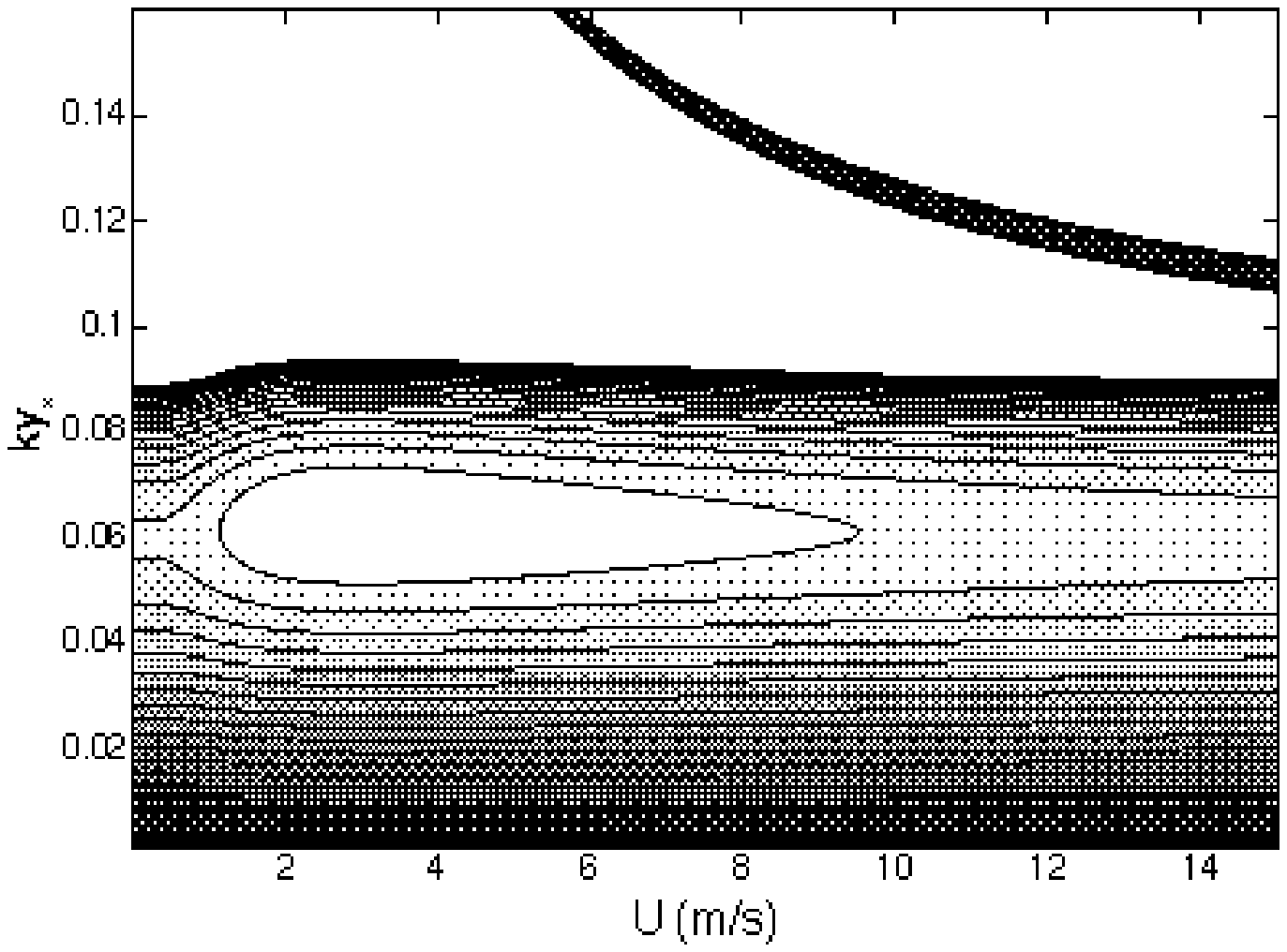} \par}

\caption{\label{fig: stab.05}The growth parameter, \protect\( \frac{\sigma _{+}v_{0}^{2}}{\left| \Gamma \right| }\protect \),
as a function of spanwise wavenumber, \protect\( ky_{*}\protect \), and background
velocity, \protect\( U\protect \), for a filament at height \protect\( v_{0}=6wu\protect \)
propagating over a wavy boundary of height to length ratio \protect\( \frac{h}{L}=0.05\protect \)
and length \protect\( L=.61mm\protect \). }
\end{figure}

\begin{figure}
{\centering \resizebox*{4.5in}{!}{\includegraphics{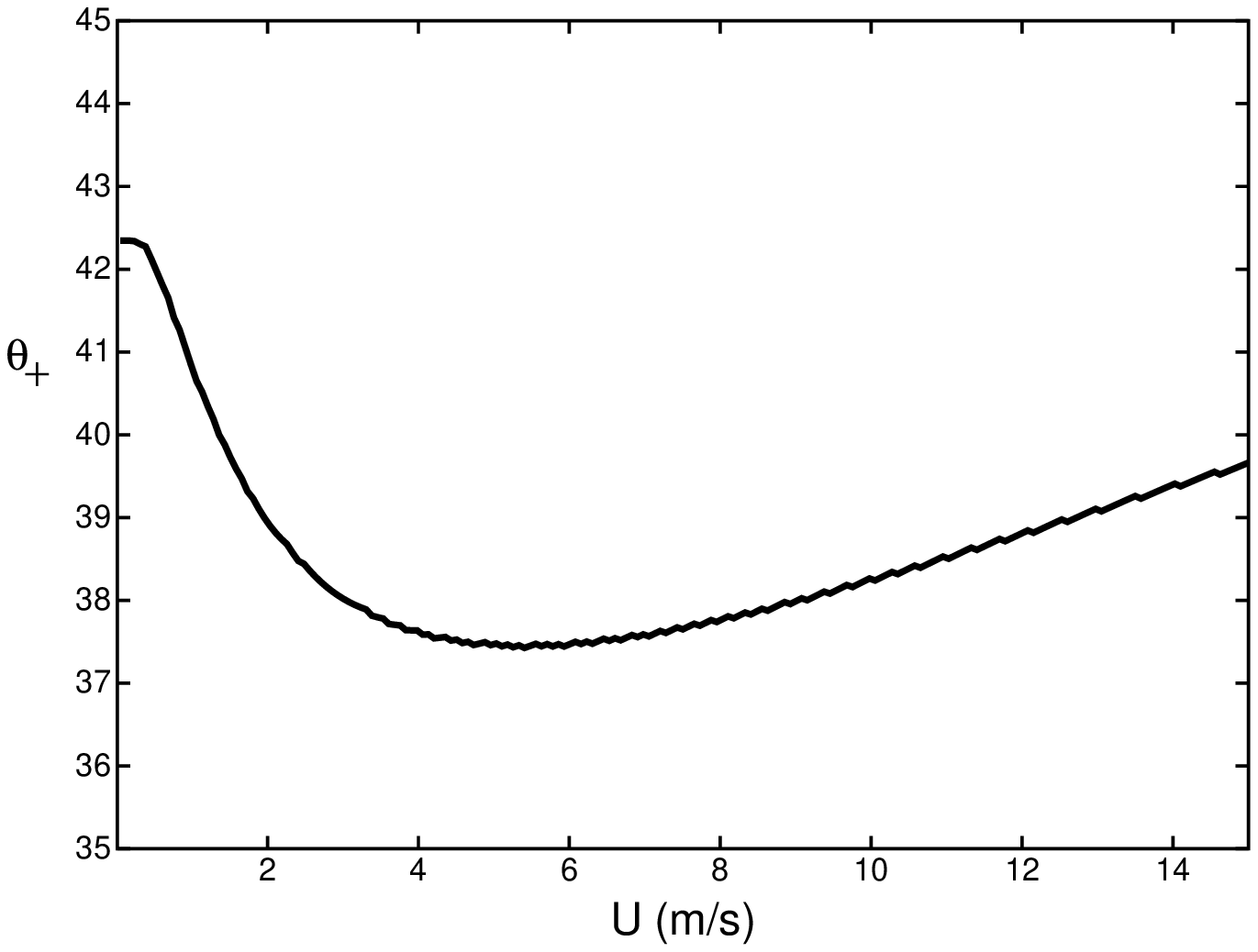}} \par}

\caption{\label{fig: staba.05}The angle of the maximally unstable mode, \protect\( \theta _{+}\protect \),
as a function of background velocity, \protect\( U\protect \), for a filament
at height \protect\( v_{0}=6wu\protect \) propagating over a wavy boundary
of height to length ratio \protect\( \frac{h}{L}=0.05\protect \) and length
\protect\( L=.61mm\protect \).}
\end{figure}

\begin{figure}
{\centering \includegraphics{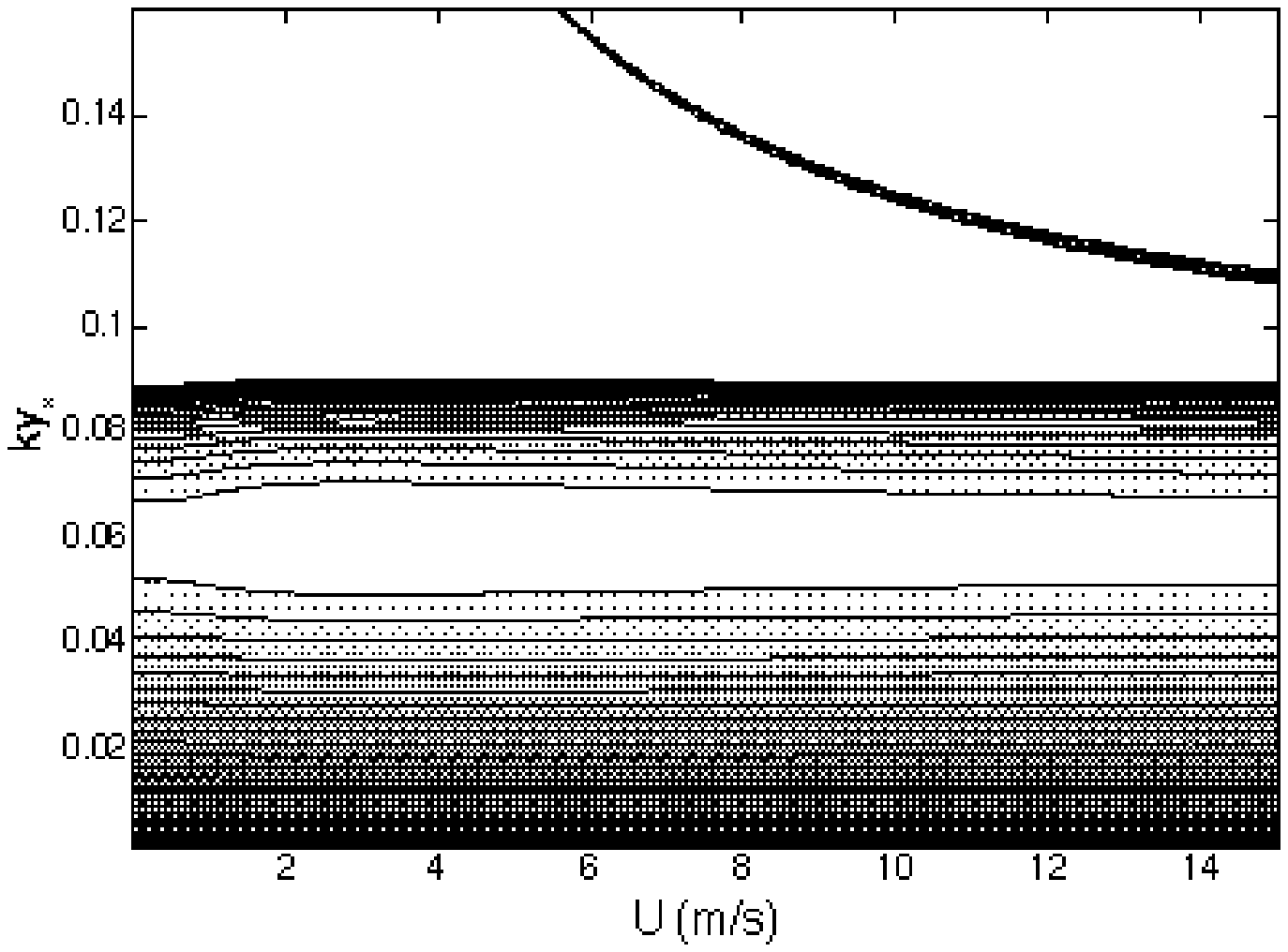} \par}

\caption{\label{fig: stab.025}The growth parameter, \protect\( \frac{\sigma _{+}v_{0}^{2}}{\left| \Gamma \right| }\protect \),
as a function of spanwise wavenumber, \protect\( ky_{*}\protect \), and background
velocity, \protect\( U\protect \), for a filament at height \protect\( v_{0}=6wu\protect \)
propagating over a wavy boundary of height to length ratio \protect\( \frac{h}{L}=0.025\protect \)
and length \protect\( L=.61mm\protect \). }
\end{figure}

\begin{figure}
{\centering \resizebox*{4.5in}{!}{\includegraphics{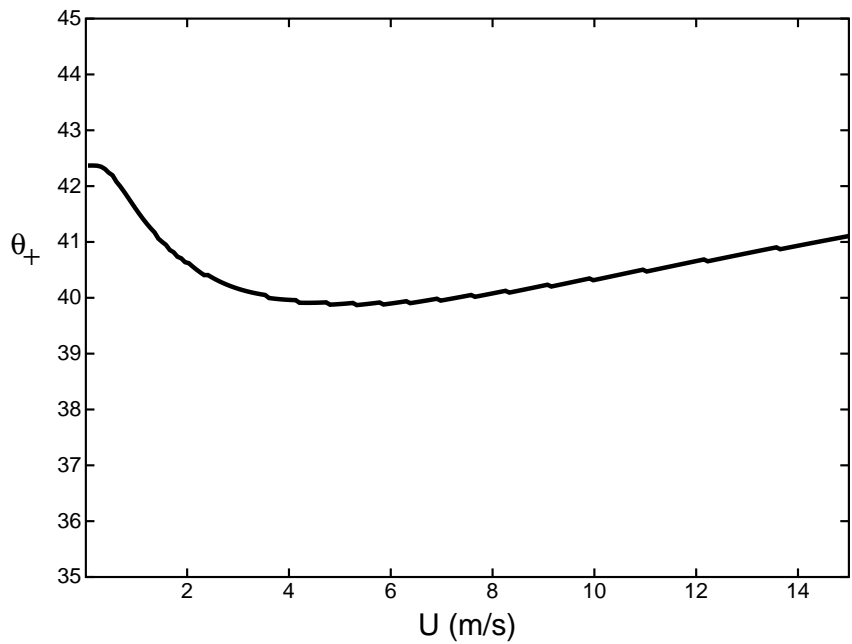}} \par}

\caption{\label{fig: staba.025}The angle of the maximally unstable mode, \protect\( \theta _{+}\protect \),
as a function of background velocity, \protect\( U\protect \), for a filament
at height \protect\( v_{0}=6wu\protect \) propagating over a wavy boundary
of height to length ratio \protect\( \frac{h}{L}=0.025\protect \) and length
\protect\( L=.61mm\protect \).}
\end{figure}

If the ridge length, \( L \), and initial filament height in wall units, \( v_{0} \),
are  taken as fixed properties of the boundary and developing flow, the instability
may be calculated for a variety of ridge heights, \( \frac{h}{L} \), over a
range of background velocities, \( U \). This gives an indication of the effect
of fixed length surface ridges for a range of swimming velocities. For delphin
ridge spacing of \( L=.61mm \) and an initial filament height of \( v_{0}=6wu \),
the instability and maximum instability angle is plotted, for ridges of height
\( \frac{h}{L}=.2,.1,.05,.025 \), over a large velocity range, (Figures \ref{fig: stab.2}-\ref{fig: staba.025}).
The maximum instability angle has a minimum at \( U\simeq 5\frac{m}{s} \) --
a velocity that is roughly independent of ridge height. The maximum instability
angle at \( U\simeq 5\frac{m}{s} \) ranges from \( 27^{o} \) for large ridges,
\( \frac{h}{L}=.2 \), to \( 40^{0} \) for very small ridges, \( \frac{h}{L}=.025 \).

\chapter{The dolphins' secret}

By altering the dynamics of vortex filaments, dolphin skin reduces the rate
of energy transport to turbulence in the evolving turbulent boundary layer.
This results in a net reduction of the drag induced by skin friction. The effectiveness
of this drag reduction technique is limited by the countering influence of increased
skin friction due to the form drag of the surface ridges. Above a critical ridge
height, \( \frac{h}{L}^{s} \), the turbulent flow over the ridges separates,
forming large eddies behind each ridge, and generates a large drag increase.
The drag reduction at a given velocity will be optimal at ridge heights just
below this critical value.

\section{The effect of surface corrugations}

The pondermotive forcing induced by the wavy boundary causes the vortex filaments
over the ridges to stretch at a maximally unstable angle, \( \theta _{+} \),
that is significantly less then the maximally unstable angle for filaments over
a flat boundary. This dynamical effect acts over a range of velocities centered
about an optimum velocity determined by the wavelength of the surface corrugations.
For a dolphin with surface ridges of wavelength \( L\simeq .61mm \), the optimal
swimming speed for this dynamic effect is \( U\simeq 5\frac{m}{s} \). Although
most effective at this velocity, surface ridges of this wavelength produce a
significant reduction in the maximally unstable angle for a range of velocities
between \( 2-15\frac{m}{s} \), (Figures \ref{fig: stab.2}-\ref{fig: staba.025}).
This result is in excellent agreement with the reported swimming speed for this
species of dolphin.

The boundary layer vortex filaments developing over such ridges grow into hairpin
vortices inclined at the reduced angle. Packets of these hairpin vortices, now
inclined at an angle lower than the \( 43^{o} \) characteristic of hairpins
over a flat surface, produce a significantly lower Reynolds stress within the
turbulent boundary layer.

\section{Implications for turbulent energy transport}

The Reynolds stress contribution from hairpin vortices inclined at an angle
\( \theta _{+} \), (\ref{rstress}), is proportional to \( \sin (2\theta _{+}) \).
Any deviation in hairpin inclination away from \( 45^{o} \) produces a reduction
of Reynolds stress, with a corresponding reduction in the rate of turbulent
energy transport. Since the dominant contribution to turbulent energy transport
comes from the hairpin vortices, and the rate of energy transport is directly
proportional to the drag force on the surface, the percentage drag reduction
due to the altered filament dynamics, \( d_{-} \), can be roughly equated with
the percentage reduction in Reynolds stress,
\[
d_{-}\simeq 100\frac{\sin \left( 2\times 43^{0}\right) -\sin \left( 2\theta _{+}\right) }{\sin \left( 2\times 43^{0}\right) }\]
The drag reduction obtained using surface ridges is directly proportional to
the degree to which the maximally unstable angle of the vortex filament may
be lowered. This angle depends upon the swimming velocity and corrugation wavelength,
as well as on the ridge height, \( \frac{h}{L} \). However, for ridge heights
that are very large, the turbulent flow over the ridges will separate, and form
large stationary eddies behind each ridge that effectively lower the ridge height
and significantly increase drag.

\section{Turbulent flow separation }

The drag reduction is limited by the maximal ridge height allowed before turbulent
flow separation is induced. A linear model of turbulent boundary layer flow
over a sinusoidal surface gives an estimate for the critical ridge height, \( \frac{h}{L}^{s} \),
at which separation occurs. This estimate is calculated, and confirmed by direct
measurement, in papers by Zilker and Hanratty \cite{dzilker1},\cite{dzilker2}.
For low friction velocities, \( Re_{L_{*}}=\frac{u_{*}L}{\nu }<630 \), and
hence for low background velocities, the critical ridge height is calculated
to be \( \frac{h}{L}^{s}\simeq .033 \). However, the critical ridge height
increases with velocity, because of the tendency of turbulent flow to inhibit
separation. For example, at \( Re_{L_{*}}\simeq 6300 \), the critical ridge
height becomes \( \frac{h}{L}^{s}\simeq .1 \). Also, because the shape of cutaneous
ridges over dolphin skin are not exactly sinusoidal, but have slightly steeper
peaks and wider troughs, the critical ridge height for separation to occur over
these ridges may vary from the value for perfectly sinusoidal ridges. It is
likely that the shape and distribution of the dolphins' cutaneous ridges is
adapted to maximize the ridge height, and the dynamical effect on the vortex
filaments, without inducing flow separation. Given the structure of cutaneous
muscle beneath the ridges, it is also very likely that the dolphins adjust the
height of their cutaneous ridges to match the swimming velocity -- keeping the
ridge height slightly below the critical value in order to maximize drag reduction.

\section{Drag reduction}

\begin{figure}
{\centering \resizebox*{4.5in}{!}{\includegraphics{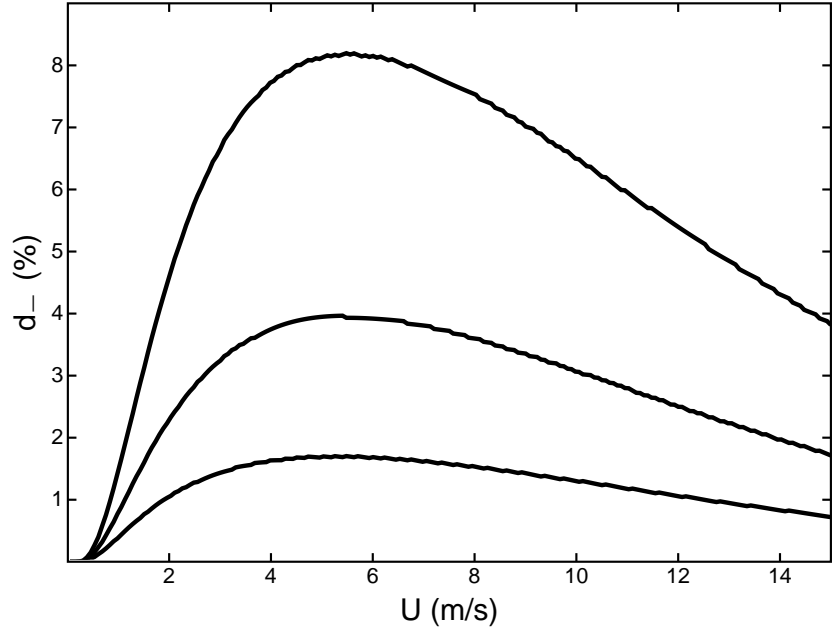}} \par}

\caption{\label{fig: drag}The estimated drag reduction due to altered filament dynamics,
\protect\( d_{-}\protect \), plotted as a function of background velocity,
\protect\( U\protect \), for wavy boundaries of length \protect\( L=.61mm\protect \)
and height to length ratio \protect\( \frac{h}{L}=.1\protect \), \protect\( \frac{h}{L}=.06\protect \),
and \protect\( \frac{h}{L}=.033\protect \).}
\end{figure}

The percent drag reduction due to the reduced inclination angle of the vortex
instability for a variety of surface ridge heights is plotted in Figure \ref{fig: drag}.
The drag reduction at the optimum swimming velocity of \( 5\frac{m}{s} \) varies
from \( 2\% \) for very small ridges of height \( \frac{h}{L}=.033 \) to \( 8\% \)
for ridges of height \( \frac{h}{L}=.1 \). Since it is likely that a dolphin
increases the ridge height with swimming velocity, the percent drag reduction
achieved is likely to increase with swimming velocity as a dolphin adjusts the
ridge height to adapt to the higher critical ridge height.

\section{Application}

Man has long looked to the animal world for inspiration in efficient engineering
design. The discovery and analysis of cutaneous ridges in odontocetes presents
another engineering feat of mother nature that may be emulated and put to human
use.

The drag reduction obtained through the use of surface corrugations will only
be significant if properly matched to the selected application. The ridge spacing,
\( L \), must be tuned for a specific velocity range, and the ridge height
must not be so great as to induce separated flow between the ridges. Although
a dolphin may vary the ridge height to accommodate greater speeds, it is unlikely
that an artificial implementation of variable ridge height would be cost effective
to manufacture and employ. Therefore, cutaneous ridges may find practical application
only when surfaces are expected to move through a fluid at a steady velocity.

The optimal ridge spacing for fluid flowing over a flat plate is calculated
to be \( L\simeq 120y_{*} \). This spacing depends only on the near boundary
structure of the turbulent boundary layer, particularly the length of one wall
unit, \( y_{*} \). The length of a wall unit over a smooth boundary may be
determined experimentally by measuring the mean surface stress, \( \tau _{0} \),
by observing the spanwise wavelength of the coherent structures present, \( \lambda _{z}\simeq 100y_{*} \),
or, alternatively, by approximating a wall unit as \( y_{*}\simeq 27\frac{\nu }{U} \).
For example, an aircraft which regularly travels at a speed of \( 150\frac{m}{s} \)
may wish to employ surface corrugations of wavelength
\[
L\simeq 120\left( 27\frac{1.5\times 10^{-5}\frac{m^{2}}{s}}{150\frac{m}{s}}\right) \simeq .3mm\]

The ridge height should be set just below the critical value, to avoid separation.
A ridge height ratio of \( \frac{h}{L}\simeq .03 \) will always be below the
critical value, and produce a surface drag reduction of \( 2\% \).  However,
the critical value of the ridge height should be determined experimentally,
so that the maximal ridge height may be employed and the greatest drag reduction
achieved. It is possible that ridge heights of ratio greater than \( .1 \)
may be allowed, producing an \( 8\% \) or higher reduction in surface drag.

\subsection{Experiment}

Although several researchers have investigated the development of a turbulent
boundary layer over a wavy wall, through experiment and numerical simulation
\cite{dzilker1}\cite{dzilker2}\cite{jhudson}\cite{vdeangelis}, none have
focused on the parameter range relevant to the flow of water over dolphin skin,
or given attention to the development and evolution of coherent structures.
However, it should be straightforward to extend these investigations to the
relevant parameter range and thus determine the validity of the theoretical
predictions.

The optimal ridge height for drag reduction, the critical height at which separation
occurs, should be determined experimentally. It would also be very usefull to
determine if a swimming dolphin does increase the cutaneous ridge height with
swimming velocity, and to what degree; although it is not clear how such an
experiment would be carried out. The dolphins have had sixty million years to
optimize the reduction of drag through the control of vortex filament dynamics,
and there is still a great deal to learn from these fascinating creatures.


\begin{thebibliography}{}
\bibitem{twill}T. Williams, ``The evolution of cost efficient swimming in marine mammals:
limits to energetic optimization,'' Phil Trans R Soc Lond B \textbf{354}, 193
(1999). 
\bibitem{jgray}J. Gray, ``Studies in animal locomotion,'' J Exp Biol \textbf{13}, 192 (1936).
 
\bibitem{mkramer}M. Kramer, ``Boundary layer control by `artificial dolphin coating','' Nav
Engnrs J \textbf{Oct}, 41 (1977).  
\bibitem{sr}S. Ridgway and D. Carder, ``Features of dolphin skin with potential hydrodynamic
importance,'' IEEE Engineering in Medicine and Biology \textbf{12}, 83 (1993).
 
\bibitem{pshoe}P. Shoemaker and S. Ridgway, ``Cutaneous ridges in odontocetes,'' Marine Mammal
Science \textbf{7}, 66 (1991).  
\bibitem{dri}``Dolphin Research Inc,'' http://www.dolphinresearch.org.au/
\bibitem{jrohr}J. Rohr, ``A novel flow visualization technique using bioluminescent marine
plankton,'' IEEE Journal of Oceanic Engineering \textbf{20}, 147 (1995).  
\bibitem{habar1}H. Abarbanel et al, ``Nonlinear analysis of high-Reynolds number flows over
a buoyant axisymetric body,'' Phys Rev E \textbf{49}, 4003 (1994). 
\bibitem{psaff}P. Saffman, \textit{Vortex Dynamics}, Cambridge University Press (1992).
\bibitem{habar2}H. Abarbanel et al, ``Vortex filament stability and boundary layer dynamics,''
Phys Rev E \textbf{50}, 1206 (1994).
\bibitem{scrow}S. Crow, ``Stability theory for a pair of trailing vortices,'' AIAA Journal
\textbf{8},2172 (1970).
\bibitem{dmoore}D. Moore, ``Finite amplitude waves on aircraft trailing vortices,'' Aeronautical
Quarterly \textbf{23}, 307 (1972).
\bibitem{jzhou}J. Zhou, R. Adrian, and S. Balachandar, ``Autogeneration of near-wall vortical
structures in channel flow,'' Physics of Fluids \textbf{8}, 288 (1996).
\bibitem{skline}S. Kline et al, ``The structure of turbulent boundary layers,'' Journal of
Fluid Mechanics \textbf{30}, 741 (1967).
\bibitem{agrass}A. Grass, ``Structural features of turbulent flow over smooth and rough boundaries,''
Journal of Fluid Mechanics \textbf{50}, 233 (1971).
\bibitem{hkim}H. Kim, S. Kline, and W. Reynolds, ``The production of turbulence near a smooth
wall in a turbulent boundary layer,'' Journal of Fluid Mechanics \textbf{50},
133 (1971).
\bibitem{mhead}M. Head and P. Banyopadhyay, ``New aspects of turbulent boundary-layer structure,''
J Fluid Mech \textbf{107}, 297 (1981).
\bibitem{ttheod}T. Theodorsen, ``Mechanisms of turbulence,'' Proc 2nd Midwestern Conf. of
Fluid Mech, Ohio State University (1952).
\bibitem{srobins}S. Robinson, ``Coherent motions in the turbulent boundary layer,'' Annu Rev
Fuid Mech \textbf{23}, 601 (1991).
\bibitem{haref}H. Aref and E. Flinchem, ``Dynamics of a vortex filament in a shear flow,''
J Fluid Mech \textbf{148}, 477 (1984).
\bibitem{arouhi}A. Rouhi and J. Wright, ``Hamiltonian formulation for the motion of vortices
in the presence of a free surface for ideal flow,'' Phys Rev E \textbf{48},1850
(1993).
\bibitem{rklein}R. Klein and O. Knio, ``Asymptototic vorticity structure and numerical simulation
of slender vortex filaments,'' J Fluid Mech \textbf{284}, 275 (1995).
\bibitem{aqi}A. Qi, ``Numerical study of wave propagation on vortex filaments,'' Journal
of Computational Physics \textbf{104}, 185 (1993).
\bibitem{jzhong}J. Zhong, T. Huang, and R. Adrian, ``Extracting 3D vortices in turbulent fluid
flow,'' IEEE Transactions on Pattern Analysis and Machine intelligence \textbf{20},
193 (1998).
\bibitem{cdelo}C. Delo and A. Smits, ``Volumetric Visualization of Coherent Structure in a
Low Reynolds Number Turbulent Boundary Layer,'' http://www.princeton.edu/\~{}gasdyn/Carl\_IJFD/DeloSmits.html
\bibitem{jbalint}J. Balint, J. Wallace, and P. Vukoslavcevic, ``The velocity and vorticity vector
fields of a turbulent boundary layer,'' J Fluid Mech \textbf{228}, 53 (1991).
\bibitem{mlevi}M. Levi and W. Weckesser, ``Stabilization of the inverted linearized pendulum
by high frequency vibrations,'' SIAM Review \textbf{37}, 219 (1995).
\bibitem{jblack}J. Blackburn, H. Smith, and N. Gronbech-Jensen, ``Stability and Hopf bifurcations
in an inverted pendulum,'' Am J. Phys \textbf{60}, 903 (1992).
\bibitem{dzilker1}D. Zilker, G. Cook, and T. Hanratty, ``Influence of the amplitude of a solid
wavy wall on a turbulent flow. Part 1. Non-separated flows,'' J Fluid Mech
\textbf{82}, 29 (1977).
\bibitem{dzilker2}D. Zilker and T. Hanratty, ``Influence of the amplitude of a solid wavy wall
on a turbulent flow. Part 2. Separated flows,'' J Fluid Mech \textbf{90},257
(1979).
\bibitem{jhudson}J. Hudson, L. Dykhno, and T. Hanratty, ``Turbulence production in flow over
a wavy wall,'' Experiments in Fluids \textbf{20}, 257 (1996).
\bibitem{vdeangelis}V. De Angelis, P. Lombardi, and S. Banerjee, ``Direct numerical simulation
of turbulent flow over a wavy wall,'' Phys. Fluids \textbf{9}, 2429 (1997).
\end{thebibliography}
\end{document}